\documentclass[journal,twoside,web]{ieeecolor}
\usepackage{generic}
\usepackage{cite}
\usepackage{amsmath,amssymb,amsfonts}
\usepackage{algorithmic}
\usepackage{graphicx}
\usepackage{algorithm,algorithmic}
\usepackage{hyperref}
\hypersetup{hidelinks=true}
\usepackage{textcomp}

\usepackage{amsmath}
\usepackage{amssymb}
\usepackage{tikz}

\usepackage{amsthm}
\usepackage{xspace}
\usepackage{array}
\usepackage{multirow}
\usepackage{cite}

\usepackage{colortbl}
\definecolor{cello}{HTML}{ffe6cc}

\usepackage{graphics} 
\usepackage{color}
\usepackage{xcolor}
\usepackage{accents}
\usepackage{centernot}
\usepackage{algorithm}
\usepackage[font=small, labelfont=bf]{caption}
\usepackage{mathtools}
\usepackage{hyperref}
\usepackage{lipsum}

\newcommand{\statedim}{n_x}
\newcommand{\localX}{\mathcal{X}_g}

\newcommand{\hjrset}{\mathcal{R}}
\newcommand{\fset}{\mathcal{S}}
\newcommand{\fseto}{\hat{\fset}}
\newcommand{\tset}{\mathcal{T}}
\newcommand{\cset}{\mathcal{U}}
\newcommand{\dset}{\mathcal{D}}
\newcommand{\sys}{f}
\newcommand{\ham}{H}
\newcommand{\maxou}{\max_{u \in \cset}}
\newcommand{\minou}{\min_{u \in \cset}}
\newcommand{\maxod}{\max_{d \in \dset}}
\newcommand{\minod}{\min_{d \in \dset}}

\newcommand{\maxerr}{\delta^*}
\newcommand{\eps}{\varepsilon}
\newcommand{\errset}{\mathcal{E}}
\newcommand{\lsys}{\ell}
\newcommand{\haml}{\ham_{\lsys}}
\newcommand{\hamlerr}{\ham_{\maxerr}}
\newcommand{\maxoe}{\max_{\eps \in \errset}}
\newcommand{\minoe}{\min_{\eps \in \errset}}

\newcommand{\infe}{\inf_{\mathfrak{e}}}
\newcommand{\infd}{\inf_{\mathfrak{d}}}
\newcommand{\supu}{\sup_{u(\cdot)}}

\newcommand{\ssp}{\mathcal{X}}
\newcommand{\ti}{t}
\newcommand{\tf}{T}
\newcommand{\tint}{[\ti,\tf]}
\newcommand{\tj}{\xi_f}
\newcommand{\csetsig}{\mathbb{U} (\ti)}
\newcommand{\dsetsig}{\mathbb{D} (\ti)}
\newcommand{\strd}{\mathfrak{d}}
\newcommand{\stratset}{\mathfrak{D}(t)}
\newcommand{\stre}{\mathfrak{e}}
\newcommand{\strateset}{\mathfrak{E}(t)}
\newcommand{\ftube}{\bar{\fset}}
\newcommand{\errsigset}{\mathbb{E} (\ti)}
\newcommand{\tjline}{\xi_{\lsys + \varepsilon}}

\newtheorem{Theorem}{Theorem}

\newtheorem{Corollary}{Corollary}

\newtheorem{Remark}{Remark}
\newtheorem*{RunningExample}{Running Example}
\def\BibTeX{{\rm B\kern-.05em{\sc i\kern-.025em b}\kern-.08em
    T\kern-.1667em\lower.7ex\hbox{E}\kern-.125emX}}
\begin{document}
\title{Conservative Linear Envelopes for Nonlinear, High-Dimensional, Hamilton-Jacobi Reachability
}


\author{William Sharpless, \IEEEmembership{Member, IEEE}, Yat Tin Chow \IEEEmembership{Member, IEEE}, and Sylvia Herbert \IEEEmembership{Member, IEEE}
\thanks{This work is supported by the UC Regents faculty fellowship, the Omnibus research grant, the NIH training grant T32 EB009380 (McCulloch), NSF DMS-2409903, ONR N000142412661 and ONR YIP N00014-22-1-2292. Sharpless and Herbert are with University of California, San Diego. Chow is with the University of California, Riverside. 
\{\href{mailto:wsharpless@ucsd.edu}{wsharpless}, \href{mailto:sherbert@ucsd.edu}{sherbert}\}@ucsd.edu, \href{mailto:yattinc@ucr.edu}{yattinc}@ucr.edu. The content is solely the responsibility of the authors.}}

\maketitle

\begin{abstract}
Hamilton-Jacobi reachability (HJR) provides a value function that encodes the set of states from which a system with bounded control inputs can reach or avoid a target despite any bounded disturbance, and the corresponding robust, optimal control policy. Though powerful, traditional methods for HJR rely on dynamic programming (DP) and suffer from exponential computation growth with respect to state dimension. The recently favored Hopf formula mitigates this ``curse of dimensionality'' by providing an efficient and pointwise approach for solving the reachability problem. However, the Hopf formula can only be applied to linear time-varying systems. 
To overcome this limitation, we show that the error between a nonlinear system and a linear model can be transformed into an adversarial bounded artificial disturbance.
One may then solve the dimension-robust generalized Hopf formula for a linear game with this ``antagonistic error" to perform guaranteed conservative reachability analysis and control synthesis of nonlinear systems; this can be done for problem formulations in which no other HJR method is both computationally feasible and guaranteed. 
In addition, we offer several technical methods for reducing conservativeness in the analysis. We demonstrate the effectiveness of our results through one illustrative example (the controlled Van der Pol system) that can be compared to standard DP, and one higher-dimensional 15D example (a 5-agent pursuit-evasion game with Dubins cars).
\end{abstract}


\section{Introduction}

\IEEEPARstart{A}{utonomous} systems across multiple domains can be described by complex nonlinear dynamics with bounded control and disturbance inputs. Applications such as urban air mobility, cellular drug delivery, and electrical grids are safety-critical, motivating the need for scalable but rigorous tools for the control design and verification of such systems. 

\begin{figure}[t]
    \centering
    \includegraphics[width=\linewidth]{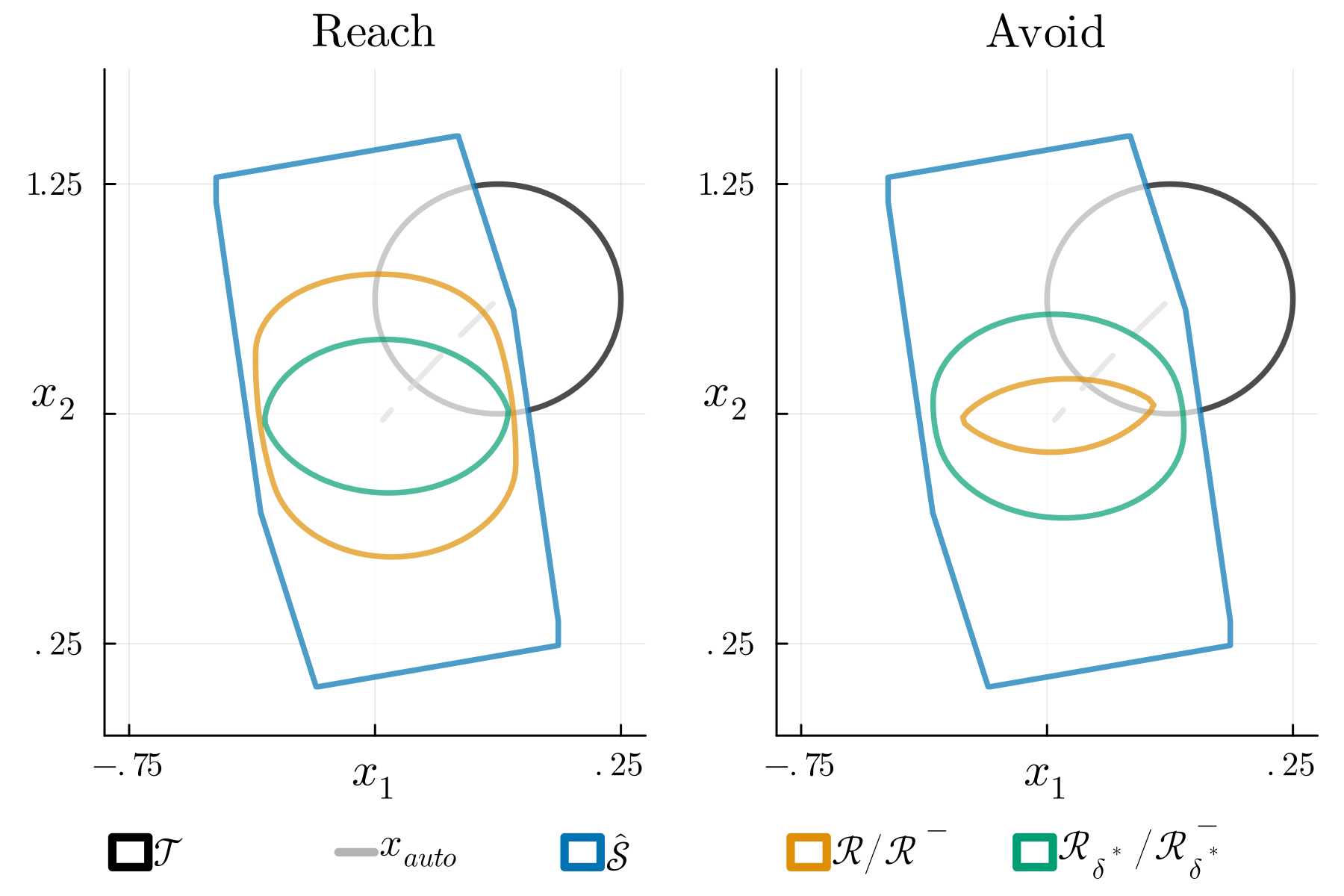}  
    \caption{\textbf{Safe Envelopes in the Van der Pol System for a Two-Player Game} The system must \textit{reach} (left) or \textit{avoid} (right) a target set $\tset$ (black). Our proposed method (green) produces guaranteed conservative reach or avoid sets $\hjrset_{\maxerr}/\hjrset^-_{\maxerr}$ compared to the true Hamilton-Jacobi reachability set $\hjrset/\hjrset^-$ (gold), solved here with dynamic-programming. The corresponding temporal slice of the conservative approximation of the feasible tube $\fseto \supset \ftube$ \cite{chen2012taylor, JuliaReach19} (blue) and the autonomous trajectory of the target center are shown (gray dash). Our proposed method is based on a linear game, here made from the Taylor series, with an antagonistic error (bounded by the maximum true error $\maxerr$ over $\fseto \supset \ftube$) and is solved by the Hopf formula, amenable to much higher dimensions than traditional Hamilton-Jacobi reachability. Thm.~\ref{thm:errenv} and Cor.~\ref{cor:saferr} prove that the resulting sets of the proposed method will be conservative and that the resulting controller will be robust to the true, nonlinear dynamics.}
    \label{fig:basic}\vspace{-2em}
\end{figure}

\textit{Hamilton-Jacobi reachability} (HJR) analysis is one of the fundamental tools for guaranteeing the safety and goal-satisfaction of a nonlinear control system. This method is based on Bellman's principle of optimality and is used to compute a value function that, first, encodes the set of states from which a nonlinear system with bounded control inputs can reach a goal (or avoid a failure set) in a given time horizon, and, second, the corresponding optimal control law that is guaranteed to achieve the goal (or avoid the failure set) despite worst-case bounded disturbance.  
This guarantee of success despite worst-case disturbances is achieved in HJR by solving a two-player zero-sum differential game, where the control seeks to minimize the distance to a goal and the disturbance instead maximizes (and vice versa for a failure set). In Figure~\ref{fig:basic}, the set represented by the black circle must be reached (left) or avoided (right) by a Van der Pol system with bounded control \eqref{vdpdynamics}. The corresponding HJR-based reach and avoid sets are shown in gold. 


The generality of the formulation and strength of the guarantees make HJR a powerful tool for robust control and verification, and it has been employed in a wide range of robotics and autonomy applications \cite{mitchell2005time, bansal2017hamilton, Kirchner_2018, herbert2017fastrack, Bokanowski10}. However, its reliance on dynamic-programming (DP) requires computation over a discretized state space. Every additional state in the dynamics model introduces exponentially more grid points for computation. Because of this, HJR is limited to 6 dimensions unless the system structure is decomposable \cite{bansal2017hamilton} or non-controllable polynomial dynamics \cite{xue2019inner}, or non-conservative, learned approximations are acceptable \cite{bansal2021deepreach, sharpless2023koopman, sharpless2024linear}.

An alternate approach to reachability is the family of \textit{differential inclusion} methods \cite{althoff2008reachability, aubin2012differential, scott2013bounds}, including zonotope propagation \cite{althoff2011zonotope} and Taylor models \cite{chen2012taylor}. These approaches are also powerful and, unlike HJR, robust to higher dimensions.  These methods over-approximate reach and avoid sets by generating a convex superset that includes all states for which  \textit{some} control and disturbance can reach the goal (or failure set) within a specified time horizon. The blue sets in Figure~\ref{fig: front_fig} demonstrate the results from one such method \cite{chen2012taylor} for the controlled Van der Pol model \eqref{vdpdynamics}. Note that for differential inclusion methods, the reach and avoid sets are identical; the analysis generally does not consider objectives but rather seeks to verify the existence of any input signals to reach the goal (or failure set). This property is also called reachability in differential inclusion works, however, in the adversarial context it is necessary but insufficient, and we distinguish it as dynamic \textit{feasibility}, i.e. there exists a pair of control and disturbance signals that will feasibly reach the goal or failure set. Generally, traditional differential inclusion methods can be overly conservative, lacking the nonlinear geometry of the exact set. The strength of these methods is that they scale to dimensions of greater than 100.

Recently, the \textit{generalized Hopf formula} was introduced as an approach to HJR without DP \cite{rublev2000generalized}. The Hopf formula solves the HJR analysis without massive spatial discretization, such that the safety value at each point may be computed independently, mitigating the curse of dimensionality. This has allowed systems of up to dimension 4096 to be solved \cite{chow2018algorithm}, however, this efficiency requires linear dynamics \cite{kurzhanski2014dynamics, darbon2016algorithms, chow2017algorithm}. Hopf reachability via linearization has been demonstrated in nonlinear systems with empirical success \cite{Kirchner_2018, sharpless2023koopman}, but guarantees are necessary for safety-critical scenarios. The authors of \cite{donggun19iterativehopf} proposed iteratively linearizing a nominal policy and adding the affine error, demonstrating convergence to a policy with improved safety, but this is not necessarily the optimal policy over space and hence may not reject the true, worst-case disturbance. The question of guaranteeing success in the nonlinear game with a linear model has remained open.

In this work, we propose an efficient, pointwise, Hopf reachability formulation for high-dimensional, nonlinear control systems. Our main results (Theorem \ref{thm:errenv} and Corollary \ref{cor:saferr}) prove that this method will generate conservative envelopes of the true HJR set and a safe controller. The key idea lies in bounding the error between a nonlinear system and a linear model of that system using differential inclusion methods, and then treating this bounded error as an adversarial disturbance to generate worst-case conservative guarantees on the resulting reach (or avoid) set and control law. 
Where the controller can overcome the antagonistic error, it can overcome the true error and thus succeeds in the true, nonlinear dynamics.
The green sets in Figure~\ref{fig: front_fig} showcase the reach and avoid sets generated by this method for the controlled Van der Pol example \eqref{vdpdynamics}. Note that they are conservative compared to the original slow HJR approach (yellow), but significantly less conservative than the differential inclusion-based approach (blue).  


We make the following novel contributions:

\begin{enumerate}

    \item We theorize a conservative, linear differential game for high-dimensional, nonlinear reachability problems. 

    \item We demonstrate several improvements for reducing the conservativeness of the guaranteed envelope.
    
    \item We outline a general, practical procedure for safely solving the problem, and demonstrate a fast-method with the Taylor series.

    \item We demonstrate the proposed work in the synthesis and verification of control policies for a 5-agent (15D) multi-agent Dubin's car model.
    
\end{enumerate}


\section{Preliminaries}\label{sec:prelims}

This paper focuses on control-affine and disturbance-affine systems of the form
\begin{eqnarray}
\dot \xi_f (\tau) \triangleq \sys(\tj, u, d, \tau)
\label{Dynamics}
\end{eqnarray}
where time $\tau \in [t,T]$, $\tj(\tau) \in \ssp \triangleq \mathbb{R}^n$, and control and disturbance inputs $u$ and $d$ are drawn from compact, convex sets $\cset \subset \mathbb{R}^{n_u}$, $\dset \subset \mathbb{R}^{n_d}$. 
W.l.o.g. we assume $T=0$.
Let the input signals $u(\cdot)$ and $d(\cdot)$ be drawn from $\csetsig  \triangleq \{ \nu:\tint \mapsto \cset \mid \nu \text{ measurable} \}$ and $\dsetsig \triangleq \{ \nu:\tint \to \dset \mid \nu \text{ measurable} \}$. Let the union of $f(x,u,d)$ over $u$ be convex and compact.
Let \eqref{Dynamics} be Lipschitz continuous in $(\tj,u,d)$ and continuous in $\tau$. For any $x \in \ssp$, let trajectory $\tj: \tint \to \ssp$ of $f$ be defined by $\tj(\ti) = x$ and $\dot{\tj}(\tau) = f(\tj(\tau), u(\tau), d(\tau), \tau)$ a.e. for $\tau \in \tint$. For clarity, we at times write $\tj(\tau; x, u(\cdot), d(\cdot), t)$.

\begin{RunningExample}
(Van der Pol) \\
To illustrate theoretical results, we consider the system 
\begin{eqnarray}
    \dot{\xi} = \begin{bmatrix} \xi_1 \\ \mu(1 - \xi_1^2)\xi_2 - \xi_1 \end{bmatrix} + \begin{bmatrix} 0 \\ 1 \end{bmatrix} (u + d) \label{vdpdynamics}
\end{eqnarray} 
with $\mu=1$, $u \in \cset \triangleq \{|u| \le 1 \}$, $d \in \dset \triangleq \{|d| \le \frac{1}{2} \}$ and a target set defined by $\tset \triangleq \{\Vert x - [0,1]^\top \Vert \le \frac{1}{4}\}$. Let $T=0$.
\end{RunningExample}

\subsection{Hamilton-Jacobi Reachability Problem}\label{sec:HJR}

To design a safe autonomous controller, HJR solves the optimal control that counters an adversarial disturbance in a differential game. 
Let the game be defined by the cost
\begin{eqnarray}
C(x, u(\cdot), d(\cdot), t) = J(\tj(T)),
\label{GameCost}
\end{eqnarray}
where $\tj(T)$ is the solution of (\ref{Dynamics}) at time $T$. 

Let the \textit{Reach} game be defined as the problem where the objective of Player I, whom chooses $u(\cdot)$, is to minimize (\ref{GameCost}) whereas the objective of Player II, whom chooses $d(\cdot)$, seeks to maximize it. Let the \textit{Avoid} game be defined as the opposite problem where Player I and Player II instead seek to maximize and minimize (\ref{GameCost}) respectively. 

Let the terminal cost $J:\ssp \mapsto \mathbb{R}$ be a convex, proper, lower semicontinuous function chosen such that
\begin{eqnarray*}
\begin{cases}
J(x) < 0 \:\: \text{ for } \:\:x \in \tset \setminus \partial\tset \\
J(x) = 0 \:\: \text{ for }\:\:x \in \partial\tset \\
J(x) > 0 \:\: \text{ for } \:\:x \notin \tset \\
\end{cases}
\label{InitialValue}
\end{eqnarray*}
where $\tset$ is a user-defined, compact set representing the target to reach or avoid and $\partial \tset$ its boundary. 
We have now defined the game(s) such that for trajectory $\tj(\cdot)$ arising from a given $x$, $u(\cdot)\in \csetsig$, $d(\cdot)\in \dsetsig$, $t$, and (\ref{Dynamics}), 
\begin{eqnarray}
C(x, u(\cdot), d(\cdot), t) \leq 0 \iff \tj(T) \in \tset.
\label{GameMeaning}
\end{eqnarray}
The functions  $V, V^-:\mathbb{R}^{n_x} \times(-\infty,T] \mapsto \mathbb{R}$ corresponding to the optimal values of the \textit{Reach} and \textit{Avoid} games respectively are defined as
\begin{eqnarray}
\begin{aligned}
V(x,t) &= \sup_{\strd} \inf_{u(\cdot)} C(x, u(\cdot), \strd[u](\cdot), t) \\
&= \sup_{\strd \in \stratset} \inf_{u(\cdot) \in \csetsig} J(\tj(T; x, u(\cdot), \strd[u](\cdot), t)), \\
V^-(x,t) &= \inf_{\strd} \sup_{u(\cdot)} C(x, u(\cdot), \strd[u](\cdot), t) \\
&= \inf_{\strd \in \stratset} \sup_{u(\cdot) \in \csetsig} J(\tj(T; x, u(\cdot), \strd[u](\cdot), t)), \\
\end{aligned}
\label{GameValue}
\end{eqnarray}
where $\stratset$ is the set of non-anticipative strategies \cite{bacsar1998dynamic, mitchell2005time},
\begin{eqnarray}
\begin{aligned}
\strd \in \stratset \triangleq \{ \gamma: \csetsig \to \dsetsig \mid u(\tau) = \hat u (\tau),\\ \text{ a.e. } \tau \in \tint \implies \gamma[u](\tau) = \gamma[\hat u] (\tau)\}.
\label{NonAntStrats}
\end{aligned}
\end{eqnarray}
Analogous to (\ref{GameMeaning}), the value functions by design have the zero-sublevel set property \cite{mitchell2005time, Bokanowski10}
\begin{eqnarray}
\begin{aligned}
V(x,t) &< 0 \iff x \in \hjrset(\tset,t), \\
V^-(x,t) &\leq 0 \iff x \in \hjrset^-(\tset,t),
\label{ValueMeaning}
\end{aligned}
\end{eqnarray}
where $\hjrset(\tset,t), \hjrset^-(\tset,t)$, are the backward reachable sets. This is the set of states which may be driven to the target despite any disturbance (\textit{Reach Set}) and despite any control (\textit{Avoid set}) respectively, formally given by
\begin{eqnarray}
\begin{aligned}
\hjrset(\tset,t) \triangleq \{x \:\:  &| \:\:  \forall \strd \in \stratset, \:\: \exists u(\cdot)\in \csetsig \\ &\text{ s.t. } \tj(t)=x, \tj (T) \in \tset \setminus \partial\tset \}, \\
\hjrset^-(\tset,t) \triangleq \{x \:\:  &| \:\:  \exists \strd \in \stratset, \:\: \forall u(\cdot)\in \csetsig \\ &\text{ s.t. } \tj(t)=x, \tj (T) \in \tset\},
\end{aligned}
\label{BRS}
\end{eqnarray}
and we assume $\tj(\cdot)$ satisfies \eqref{Dynamics} s.t. $\dot{\tj}(\tau) = f(\tj(\tau), u(\tau), \frak{d}[u](\tau), \tau)$ a.e. on $[t,T]$. In contrast, consider the set of dynamically \textit{feasible} states $\fset$ for which there exists inputs in $\cset$ and $\dset$ that \textit{could} arrive at the target, given by
\begin{eqnarray}
\begin{aligned}
\fset(\tset, t) \triangleq \{x \:\:  |& \:\:  \exists u(\cdot)\in \csetsig \:\: \exists d(\cdot)\in \dsetsig \\ &\text{ s.t. } \tj(t)=x, \tj(T) \in \tset\}.
\label{rsetdef}
\end{aligned}
\end{eqnarray}
Given the compactness of $(\tset \times \cset \times \dset \times [t,T])$, the measureability of $u(\cdot)$ and $d(\cdot)$, and the Lip. continuity of $\sys$, by Filipov and others \cite{filippov1960differential, aubin2012differential, scott2013bounds} we may know the existence and boundedness of this set. The feasible tube $\ftube$ is given by
\begin{eqnarray}
\begin{aligned}
\ftube(\tset, t) \triangleq \bigcup_{\tau \in \tint} \fset(\tset,\tau),
\label{ftubedef}
\end{aligned}
\end{eqnarray}
which we may also know is bounded given the above assumptions for the closed interval $\tint$ \cite{aubin2012differential, scott2013bounds}. 
When considering antagonistic or worst-case scenarios, this \textit{feasibility} is insufficient to guarantee $\tj(T) \in \tset$. However, $\fset \supset \hjrset, \hjrset^-$ for any $(\tset, t)$, and, in particular, one may rapidly over-approximate $\fset$ \& $\ftube$ with convex differential inclusions \cite{althoff2008reachability, aubin2012differential, chen2012taylor}, thus, they serve as valuable objects for bounding the relevant errors. 

Notably, applying Bellman's principle of optimality to the value function $V$ leads to the following well-known theorem. 

\begin{Theorem}
[Evans 84] \cite{evans1984differential} \\
The value function $V$ defined in (\ref{GameValue}) is the viscosity solution to the following HJ-PDE,
\begin{eqnarray}
\begin{aligned}
\dot{V}  + \ham(x, \nabla V, \tau) & = 0 &\text{ on } \ssp& \times [t,T], \\
V(x,T) & = J(\tj(T)) &\text{ on } \ssp&, 
\end{aligned}
\label{HJPDE-V}
\end{eqnarray}
where the Hamiltonian $\ham:\ssp \times \ssp \times [t,T] \mapsto \mathbb{R}$ is 
\begin{eqnarray}
\ham(x, p, \tau) = \minou \maxod p \cdot \sys(x, u, d, \tau).
\label{Hamiltonian}
\end{eqnarray}
\label{thm:HJPDE}
\vspace{-1em}
\end{Theorem}
This theorem equivalently applies to $V^-$, and the Hamiltonian in the \textit{Avoid} game takes the flipped form $H^-(x,p,\tau)=\max_u \min_d p \cdot f(x,u,d,\tau)$. To solve this PDE, therefore, yields the value function and corresponding reachable set. Additionally, the value functions can be used to derive the optimal control strategy for any point in space and time, e.g., for the \textit{Reach} game, for any $x \in \ssp$ and $\tj(t) = x$, $\forall \tau \in \tint$
\begin{eqnarray}
\begin{aligned}
u^*(\tau) \in \arg \minou \maxod \nabla V(\tj(\tau),\tau) \cdot f(\tj(\tau), u, d, \tau).
\end{aligned}
\label{HJoc}
\end{eqnarray}
The main challenge of HJR lies in solving the PDE in (\ref{HJPDE-V}); DP-based methods propagate $V(x, t)$ by finite-differences over a grid of points that grows exponentially with respect to $\statedim$ \cite{bansal2017hamilton}. In practice, this is intractable for systems of $\statedim \geq 6$.

\subsection{The Hopf Solution to HJ-PDE's}\label{subsec:hopf}

An alternative to the DP-based HJR method for computing $V(x,t)$ or $V^-(x,t)$ is the Hopf formula, which offers a solution to (\ref{HJPDE-V}) in the form of a pointwise optimization problem. This avoids discretization over a grid and, thus, permits rapid and efficiently feedback for control.  Note, we omit the Hopf analysis of the \textit{Avoid} game in this section for brevity; however, it may be derived identically.
Let $\underline{t} \triangleq T-t$ and $\underline{\tau} \in [0, \underline{t}]$, then we may define $\phi(x,\underline{t}) \triangleq V(x, t)$ to simplify the following equations. Thus, $\phi$ is the solution of
\begin{eqnarray}
\begin{aligned}
- \dot{\phi} + \ham(x, \nabla \phi, \underline{\tau}) &= 0 &\text{ on } \ssp& \times [0,\underline{t}], \\
\phi(x,0) &= J(x) &\text{ on } \ssp&, 
\end{aligned}
\label{HJPDE-phi}
\end{eqnarray}
with Hamiltonian,
\begin{eqnarray}
\ham(x, p, \underline{\tau}) = \maxou \minod -p \cdot \sys(x, u, d, T-\underline{\tau}).
\label{HamiltonianPhi}
\end{eqnarray}
Note, for systems $\sys(x,u,d,\tau)=\sys(u,d,\tau)$ without state dependence, the Hamiltonian $\ham(x,p,\underline{\tau}) = \ham(p,\underline{\tau})$ also lacks state-dependence and in this setting the following Hopf formula is available with limitation. This formula was conjectured in \cite{hopf1965generalized}, proved to be the viscosity solution in \cite{bardi1984hopf} and \cite{lions1986hopf} for $\ham(p)$, and generalized to time-varying $\ham(p,\underline{\tau})$ in \cite{rublev2000generalized, kurzhanski2014dynamics}. Recently, \cite{darbon2016algorithms} devised a fast method of solving this formula and \cite{chow2019algorithm} conjectured a general form for $\ham(x,p,\underline{\tau})$.

\begin{Theorem}
[Rublev 00] \label{thm:rublev} \cite{rublev2000generalized} \\
Assume that $J(x)$ is convex and Lipschitz, and that $\ham(p,\underline{\tau})$ is pseudoconvex in $p$ and satisfies (B.i-B.iii) in [Rublev 00], then the minimax-viscosity solution \cite{subbotin1996minimax} of (\ref{HJPDE-phi}) is given by the time-dependent Hopf formula
\begin{eqnarray}
\phi(x,\underline{t}) = -\min_{p \in \mathbb{R}^{\statedim}} \bigg\{ J^\star(p) - x \cdot p + \int_0^{\underline{t}} \ham(p, \underline{\tau}) d\underline{\tau} \bigg\}
\label{HopfFormula}
\end{eqnarray}
where $J^\star(p):\ssp \rightarrow \mathbb{R} \cup \{+\infty\}$ is the Fenchel-Legendre transform (i.e. convex-conjugate) of a convex, proper, lower semicontinuous function $J:\ssp \rightarrow \mathbb{R}$ defined by
\begin{eqnarray}
J^\star(p) = \sup_{x\in \ssp} \{ p \cdot x - J(x) \}.
\label{FL}
\end{eqnarray}
\end{Theorem}

\begin{Remark}
Under these assumptions, the minimax-viscosity and viscosity solutions are equivalent when $\ham(p, \underline{\tau})$ is convex or concave in $p$ for $\underline{\tau} \in [0,\underline{t}]$ \cite{rublev2000generalized, subbotin1996minimax}.
\end{Remark}

See \cite{rublev2000generalized, subbotin1996minimax, chow2019algorithm} for analysis and comparison of minimax-viscosity and viscosity solutions. When the solutions differ, only the latter corresponds to the solution of the differential game, thus, one prefers a convex Hamiltonian for a \textit{Reach} game in practice. For general non-convex $\ham$, the question of when these solutions coincide remains open, however, some guarantees still exist for certain problem formulations as noted in Section~\ref{sec:SfEnv} and Remark \ref{rem:ncHavoid}.

The strength of the Hopf formula is that \eqref{HJPDE-V} and \eqref{HJPDE-phi} may be computed by solving a \textit{pointwise} optimization problem \cite{darbon2016algorithms, chow2017algorithm, chow2019algorithm}, mitigating the \textit{curse of dimensionality} compared to DP-based methods. 
However, the formulation in \eqref{HopfFormula} makes the strong assumption of state independence of the Hamiltonian (and therefore also assumes state independence in the system dynamics). 
For linear time-varying systems, this assumption can be managed by mapping the system to a state-independent form \cite{kurzhanski2014dynamics}, e.g. 
\begin{eqnarray}
\begin{aligned}
\dot x = A(\tau)x + &B_1(\tau) u + B_2(\tau) d \\ \rightarrow \dot z = \Phi(\tau) (&B_1(\tau) u + B_2(\tau) d),
\end{aligned}
\label{LinearStateIndependent}
\end{eqnarray}
with the linear time-varying mapping $z \triangleq \Phi(\tau) x$, where $\Phi(\tau)$ is the matrix defined for $\tau \in [t, T]$ by
\begin{equation}
\dot \Phi = - \Phi(\tau) A(\tau), \quad \Phi(T) = I.
\end{equation}
After the changes of variables, the state-independent Hamiltonian for $\mathcal{Z}$ becomes,
\begin{eqnarray}
\begin{aligned}
\ham_\mathcal{Z} (p, \underline{\tau}) =& \maxou -p \cdot \Phi(T-\underline{\tau}) B_1(T-\underline{\tau}) u \\ 
& - \maxod -p \cdot \Phi(T-\underline{\tau}) B_2(T-\underline{\tau}) d .
\end{aligned}
\label{HamiltonianLinear}
\end{eqnarray}
Since the mapping $\Phi$ is injective, we may define $\phi_\mathcal{Z}(z, \underline{t})$ by
\begin{eqnarray}
\phi_\mathcal{Z}(z, \underline{t}) \triangleq -\min_{p} \bigg\{ J_\mathcal{Z}^\star(p) - z \cdot p + \int_0^{\underline{t}} \ham_\mathcal{Z}(p, \underline{\tau}) d\underline{\tau} \bigg\}
\label{HopfFormulaZ}
\end{eqnarray}
where $J_\mathcal{Z}(z) \triangleq J([\Phi(T - \underline{\tau})]^{-1} z)$ s.t. 
$J_\mathcal{Z}^\star(p) = J^\star(p)$ 
and ultimately $\phi_\mathcal{Z}(z, \underline{t}) = \phi(x, \underline{t})$ (see \cite{chow2017algorithm, donggun19iterativehopf} for details).

Given the convexity of $\cset$ and $\dset$, the Hamiltonian in (\ref{HamiltonianLinear}) can be rewritten as the difference of two positively homogeneous Hamiltonians corresponding to the convex-conjugates of their indicator functions (\ref{Indicator})  \cite{chow2017algorithm},
\begin{eqnarray}
\begin{aligned}
\ham_\mathcal{Z} (p,\underline{\tau}) = \mathcal{I}_\cset^\star(R_1(\underline{\tau}) p) - \mathcal{I}_\dset^\star(R_2(\underline{\tau}) p), \\ R_i(\underline{\tau})\triangleq-B_i(T-\underline{\tau})^\top \Phi(T-\underline{\tau})^{\top} .
\end{aligned}
\label{HamiltonianIndicator} 
\end{eqnarray}
where
\begin{eqnarray}
\mathcal{I}_\mathcal{B}(b) \triangleq \{0 \text{ if } b \in \mathcal{B}, +\infty \text{ else} \}.
\label{Indicator}
\end{eqnarray}
Convexity of the Hamiltonian may be shown with this relationship depending on the forms of $\cset$ \& $\dset$, and if convex, it is said the control has more \textit{authority} than disturbance (refer to \cite{sharpless2023koopman} for more details). One should note that in the \textit{Avoid} game with $V^-$, for $H^-$ to be convex then the difference is flipped, i.e. the disturbance must have more authority. However, there is an interesting asymmetry between the \textit{Reach} and \textit{Avoid} game with the Hopf formula.

\begin{Remark} (\textit{Avoid} Games may have Non-Convex $\ham$) \label{rem:ncHavoid} \\
    The minimax-viscosity solution given by the Hopf formula is always less than or equal to the viscosity solution \cite{wei2013viscosity}. Hence, in an avoid problem,  $\phi^-$ (the Avoid Hopf value) guarantees a safe lower bound of $V^-$ for any $H^-$ satisfying Thm.~\ref{thm:rublev}. \footnote{This yields a potentially non-convex objective, thus one should employ ADMM, known to solve similar non-convex problems \cite{wang2019global}.} 
\end{Remark}

\section{Safe Envelopes with Linear Systems}\label{sec:SfEnv}


In this section, we propose a method to bound the error from any linearization and to treat this error as an adversarial player in the differential game to guarantee the linear solution will be conservative in the reachability analysis.

\subsection{Antagonistic Error Formulation}\label{subsec:AntagErr}
Consider a continuous, linear function $\lsys$ that will serve as an approximation of the dynamics \eqref{Dynamics} with the form 
\begin{eqnarray}
\lsys (x,u,d,\tau) \triangleq A(\tau)x + B_1(\tau) u + B_2(\tau) d + c.
\label{LinearSystem}
\end{eqnarray}

In the entire domain $\mathcal{X}$, the error for $\lsys$ may be unbounded. Instead, consider the feasible tube $\ftube$ of all states that \textit{could} travel into the target for some control and disturbance, defined in (\ref{ftubedef}). 
Given the boundedness of this set which arises from the previous assumptions, the maximum error on $\ftube$ between the nonlinear system \eqref{Dynamics} and linear system \eqref{LinearSystem},
\begin{eqnarray}
\maxerr \triangleq \max_{\ftube(\tset, t) \times \Sigma (t)} \left\Vert \big[ \sys - \lsys \big](x, u, d, \tau) \right\Vert,
\label{maxerrdeftube}
\end{eqnarray}
is finite, where $\Sigma (t) \triangleq \cset \times \dset \times \tint$. Then for $x \in \ftube(\tset, t)$, we may rewrite system (\ref{Dynamics}),
\begin{eqnarray}
\dot{x} = \sys(x,u,d, \tau) = \lsys(x,u,d, \tau) + \eps, \quad \Vert\eps\Vert \le \maxerr.
\label{LinearEpsilon}
\end{eqnarray}
Moreover, the Hamiltonians may be rewritten, 
\begin{eqnarray}
\begin{aligned}
H(x, p, \tau) = \haml(x,p,\tau) + p \cdot \eps, \\
H^-(x, p, \tau) = \haml^-(x,p,\tau) + p \cdot \eps,
\label{LinearHamiltonianEps}
\end{aligned}
\end{eqnarray}
where $\haml, \haml^-$ are the Hamiltonians of the linear system, e.g. $\haml(x,p,\tau) \triangleq \minou \maxod p \cdot \lsys(x, u, d, \tau)$.
Although $\eps$ is a nonlinear, state-dependent quantity, we may assume the worst-case value of $\eps$ in the analysis to be generate safe results.

Allow an error player to act antagonistically toward the control (in league with the disturbance), drawing from the error set $\errset \triangleq\{ \eps \in \mathcal{X} \mid \Vert \eps \Vert \le \maxerr \}$. Let $\errsigset$, with $\eps(\cdot):\tint \mapsto \errset$, be the measurable error signals and $\strateset$, with $\frak{e}:\csetsig \mapsto \errsigset$, be the non-anticipative error strategies (defined analogous to $\dsetsig$ and $\stratset$ in Sec.~\ref{sec:HJR}). Then the corresponding values for the \textit{Reach} and \textit{Avoid} games with dynamics $\lsys$ and antagonistic error take the form,
\begin{equation*}
\begin{aligned}
   &V_{\maxerr}(x,t) \triangleq \\ &\sup_{\stre \in \strateset} \sup_{\strd \in \stratset} \inf_{u(\cdot) \in \csetsig} J(\tjline (T; x, u(\cdot), \strd[u](\cdot), \stre[u](\cdot), t)), \\
   &V_{\maxerr}^-(x,t) \triangleq \\ & \inf_{\stre \in \strateset} \inf_{\strd \in \stratset} \sup_{u(\cdot) \in \csetsig} J(\tjline (T; x, u(\cdot), \strd[u](\cdot), \stre[u](\cdot), t)),
\end{aligned}
\end{equation*}
where $\tjline$ are trajectories of \eqref{LinearEpsilon} with inputs: $u \in \cset, d \in \dset, \eps \in \errset$. Moreover, since $\errset$ is compact, it follows from Thm.~\ref{thm:HJPDE} that $V_{\maxerr}(x,t)$ and $V_{\maxerr}^-(x,t)$ are viscosity solutions of the HJB PDE \cite{evans1984differential} in (\ref{HJPDE-V}) with Hamiltonians,
\begin{eqnarray}
\begin{aligned}
\hamlerr(x, p, \tau) \triangleq \haml(x,p,\tau) + \maxoe p \cdot \eps, \\
\hamlerr^-(x, p, \tau) \triangleq \haml^-(x,p,\tau) +  \minoe p \cdot \eps,
\label{LinearHamiltonianEps}
\end{aligned}
\end{eqnarray}
In comparison with \eqref{LinearHamiltonianEps}, one may observe $\hamlerr=\maxoe \ham$ and hence $\hamlerr$ is the \textit{envelope} of the true Hamiltonian in \eqref{Hamiltonian} in the parlance of PDEs \cite{evans2014envelopes}. Note, $\ftube$ with $\maxerr$ is viable to make the linear game with antagonistic error conservative, but a more general condition holds.
We now consider the principal result of the work.

\begin{Theorem}  \label{thm:errenv}
    Let any bounded set $\fseto_{c}$ be s.t. for some $c \in \mathbb{R}$,
    \begin{eqnarray}
    \begin{aligned}
    \fseto_{c} \supseteq \{ y \in \ssp  \mid  \:\: &y = \tj(\tau; x, u(\cdot), d(\cdot), t), \text{ s.t. } \\ &J(\tj(T)) \le c, \tau \in \tint\},
    \label{errenvhypo}
    \end{aligned}
    \end{eqnarray}
   Let $\maxerr \triangleq \max_{\fseto_{c} \times \Sigma (t)} \Vert [f - \lsys] (x, u, d, \tau) \Vert$. Then for any $x \in \mathcal{V}_{c}$, the $c$-level set of $V$, and any $t$,
    \begin{align}
       & V (x,t) \le V_{\maxerr} (x,t), \label{eq:ConservativeReachVal} 
       \quad V^- (x,t) \ge V_{\maxerr}^- (x,t). 
    \end{align}
\end{Theorem}

Before presenting the proof, note that the bounded set $\fseto_{c}$ contains 
all trajectories evolving from $x$ into the $c$-sublevel set of $J$. It follows that if the antagonistic error action is bounded by the maximum error $\maxerr$ on this set, then there exists a strategy for the error player $\tilde \stre \in \strateset$ which induces the true trajectory $\tj$ and may be played to their benefit. Hence, the supremum or infimum over all error strategies will be conservative with respect to the true game.

There are several options for what $\fseto_{c}$ may be (see Cors.~\ref{cor:spFS}, \ref{lem:par}, \ref{lem:FBerr}), the most straightforward of which is the feasible tube $\ftube$ (Cor. \ref{cor:saferr}) which may be computed rapidly with differential inclusions (Sec. \ref{sec:det}). 
As we will show in the subsequent corollaries, together these claims will lead to conclusions on the conservativeness of the backward reachable sets and the controllers generated from the linear game with antagonistic error. The proof of Thm.~\ref{thm:errenv} follows.
\begin{proof}
    Both \textit{Reach} and \textit{Avoid} proofs follow from construction of a specific strategy for the antagonistic error; we show the \textit{Avoid} only. Namely, we prove the following contrapositive statement that, for any $\bar c \le c$, we have $$V^- (x,t) < \bar c \implies V_{\maxerr}^- (x,t) < \bar c.$$

    Consider $\bar c \le c$. We realize
    \begin{eqnarray}
    \begin{aligned}
    \fseto_{c} \supseteq \{ y \in \ssp  \mid  \:\: &y = \tj(\tau; x, u(\cdot), d(\cdot), t), \text{ s.t. } \\ &J(\tj(T)) \le \bar c, \tau \in \tint\}.
    \end{aligned}
    \end{eqnarray}
    Note, $x \notin \mathcal{V}_{c}$ trivially gives the RHS of \eqref{eq:ConservativeReachVal}, hence we need only consider $x \in \mathcal{V}_{c} \subseteq \fseto_{c}$. Assume that 
    $$ \bar c > V^- (x,t) = \infd \supu  J(\tj(T; x, u(\cdot), \strd[u](\cdot), t)).$$
    For time $\tau \in \tint$, let the error for trajectory $\tj$ be
    \begin{equation}
    \begin{aligned}
        \varepsilon(\tau; &\: \tj, u(\cdot), d(\cdot)) \triangleq \\ &f(\tj(\tau), u(\tau), d(\tau)) - \lsys(\tj(\tau), u(\tau), d(\tau)).
        \label{errdef}
        \end{aligned}
    \end{equation}
    Then we may transform trajectories $\tj$ into trajectories of the linear system $\lsys$ with error $\varepsilon$, say $\tjline$. For proof, consider their difference at time $s\in\tint$,
    \begin{eqnarray*}
    \begin{aligned}
        &\Vert \tj(s) - \tjline(s) \Vert \\
        &= \bigg \Vert x + \int_t^s f(\tj(\tau), u(\tau), d(\tau))\, d\tau \\ &\qquad - \left(x + \int_t^s \lsys(\tjline(\tau), u(\tau), d(\tau)) + \varepsilon(\tau) \, d\tau \right) \bigg \Vert \\
        &= \left \Vert \int_t^s \lsys(\tj(\tau), u(\tau), d(\tau)) - \lsys(\tjline(\tau), u(\tau), d(\tau)) \, d\tau \right \Vert \\ &\le L_\lsys \int_t^s \Vert \tj(\tau) - \tjline(\tau) \Vert \, d\tau .
    \end{aligned}
    \end{eqnarray*}
    where $L_\lsys$ is the Lipschitz constant for $f$. Writing $\phi(\tau) = \int_t^\tau \Vert \tj (s) - \tjline (s) \Vert \, ds$, then we directly have $\dot \phi - L_\lsys \phi \le 0, \phi \ge 0, \phi(t) = 0$, and the Gronwall inequality gives $ 0 \le \phi(\tau) \le \phi(t) \exp( L_\lsys t ) = 0 $, thus $\phi(\tau) \equiv 0$ \cite{boyce2021elementary}. Hence, $\tj(\tau) = \tjline(\tau), \forall \tau \in [t,T]$. 
    
    Then, from the original assumption,
    $$ \bar c > \infd \supu J(\tjline (T; x, u(\cdot), \strd[u](\cdot), \varepsilon(\cdot), t)).$$
    $\implies  \exists \sigma > 0, \exists \tilde \strd \in \stratset$
    \begin{equation}
        \bar c > \bar c-2\sigma > \supu J(\tjline (T; x, u(\cdot), \tilde \strd[u](\cdot), \varepsilon(\cdot), t), \label{supJ}
    \end{equation}
    $\implies  \exists \sigma > 0, \exists \tilde \strd \in \stratset, \forall u(\cdot) \in \csetsig$
    \begin{equation}
        \bar c > \bar c- \sigma > J(\tjline (T; x, u(\cdot), \tilde \strd[u](\cdot), \varepsilon(\cdot), t).
        \label{inJc}
    \end{equation}
    Let $\tilde \stre$ be a strategy s.t. $\forall u \in \csetsig$, 
    $$\tilde \stre[u](\cdot) \triangleq \varepsilon(\cdot; \tj(\cdot; x, u(\cdot), \tilde \strd[u](\cdot), t), u(\cdot), \tilde \strd[u](\cdot)).$$ 
    Note, by (\ref{inJc}) and the definition of $\fseto_{c}$ and $\maxerr$, then $\forall \tau, \tilde \stre[u](\tau) \in \errset$. Moreover, since $u(\cdot)$ \& $ \tilde \strd[u](\cdot)$ are measurable and $f$ \& $\lsys$ are Lipschitz, then from (\ref{errdef}) $\tilde \stre[u](\cdot)$ must be measurable, and $\tilde \stre[u](\cdot) \in \errsigset$. If $u(\tau) = \hat u (\tau)$, then since $\tilde \strd$ is non-anticipative, $\tilde \strd[u](\tau) = \tilde \strd[\hat u](\tau)$, and with (\ref{errdef}), $\tilde \stre[u](\tau) = \tilde \stre[\hat u](\tau)$ and $\tilde \stre \in \strateset$.
    
    Then from (\ref{supJ}),
    \begin{align*}
    \bar c > &\supu J(\tjline (T; x, u(\cdot), \tilde \strd[u](\cdot), \tilde \stre[u](\cdot), t), \\
    \implies \bar c > &\infe \infd \supu J(\tjline (T; x, u(\cdot), \strd[u](\cdot), \stre[u](\cdot), t))\\ 
    & = V_{\maxerr}^- (x,t).
    \qedhere
    \end{align*}
\end{proof}
Therefore, we may know that with a sufficient $\fseto_{c}$ then we may have a conservative solution w.r.t. the true value, implying the following more practical result.

\begin{Corollary} \label{cor:saferr} 
In the \textit{Reach} and \textit{Avoid} games, if $V_{\maxerr}$ and $V_{\maxerr}^-$ are defined with $\maxerr$ in (\ref{maxerrdeftube}), then $\forall \tau \in [t,T]$,
\begin{align}
        \hjrset_{\maxerr}(\tset,\tau) &\subset \hjrset(\tset,\tau) \label{eq:ConservativeReachSet} \\
        \hjrset_{\maxerr}^-(\tset, \tau) &\supset \hjrset^-(\tset, \tau), \label{eq:ConservativeRAvoidSet}
\end{align}
Moreover, for any $x \in \hjrset_{\maxerr}(\tset,\tau)$ or $x \in \hjrset^-_{\maxerr}(\tset,\tau)$, $u^*_{\maxerr}(\cdot)$ solved from (\ref{HJoc}) with $V_{\maxerr}$ or $V_{\maxerr}^-$ will yield $\tj(T) \in \tset$ or $\tj(T) \notin \tset$ respectively under the true, nonlinear dynamics (\ref{Dynamics}) despite any disturbances $d(\cdot)\in \dsetsig$.
\end{Corollary}
\begin{proof} The \textit{Reach} and \textit{Avoid} proofs are similar, hence, only the \textit{Reach} is shown for brevity. In Thm.~\ref{thm:errenv}, let $\bar c=0$ and then by definition (\ref{ftubedef}), $\ftube(\tset, t)$ satisfies the hypothesis (\ref{errenvhypo}). It then follows from the game value property (\ref{ValueMeaning}),
\begin{eqnarray*}
\begin{aligned}
x \in \hjrset_{\maxerr}(\tset,\tau) \iff &V_{\maxerr}(x,\tau) \le 0 \\
\implies &V(x,\tau) \le 0 \iff x \in \hjrset(\tset,\tau).
\end{aligned}
\end{eqnarray*}
By definition, $x \in \hjrset_{\maxerr}(\tset,\tau)$ implies
\begin{eqnarray*}
 \forall \stre \in \strateset, \forall \strd \in \stratset, \exists u^*_{\maxerr}(\cdot) \in \csetsig, \text{ s.t.} \\
 \tjline(T; x, u^*_{\maxerr}(\cdot), \strd[u^*_{\maxerr}](\cdot), \stre[u^*_{\maxerr}](\cdot), t) \in \tset, 
\end{eqnarray*}
which is given by \eqref{HJoc} with $V_\maxerr$ \cite{evans1984differential}. Finally, by the choice of $\maxerr$, $\exists \tilde \stre \in \strateset$ s.t. $\tjline(\cdot; u^*_{\maxerr}(\cdot), \strd[u^*_{\maxerr}](\cdot), \tilde \stre[u^*_{\maxerr}](\cdot)) = \tj(\cdot; u^*_{\maxerr}(\cdot), \strd[u^*_{\maxerr}](\cdot)) \implies \tj(T; u^*_{\maxerr}(\cdot), \strd[u^*_{\maxerr}](\cdot)) \in \tset \qedhere$.   

\end{proof}

Thm.~\ref{thm:errenv} and Cor.~\ref{cor:saferr} certify that solving the original games with the linear dynamics and antagonistic error (\ref{LinearEpsilon}) yields a conservative subset (superset) of the true Reach (Avoid) Set, and the optimal control computed from this game is impervious to the true nonlinear dynamics. This is crucial because the linear system with error given in (\ref{LinearEpsilon}) may be solved with the dimension-robust Hopf formula, namely via the z-transformation \eqref{HopfFormulaZ} of $\hamlerr$ and $\hamlerr^-$.
This is demonstrated in the running example in the Van der Pol system in Fig.~\ref{fig:basic}. 

While this is applicable to a general class of systems and games, a large nonlinearity will yield an over-conservative set. Namely, for a local linearization with an error that increases with distance, these safe envelopes will thus have limited spatial efficacy. This can also corrupt the convexity requirement of the Hopf formula. To improve this, we propose a time-varying extension for greatly decreasing the error player's authority and thus offering a tighter envelope. We offer several other extensions for practical use in the Supplementary (see \href{https://drive.google.com/file/d/13gXlDd5__ZoE15QeyZeEQm3pa_vlUxzI/view?usp=sharing}{online}).



\begin{figure}[t]
    \centering
    \includegraphics[width=\linewidth]{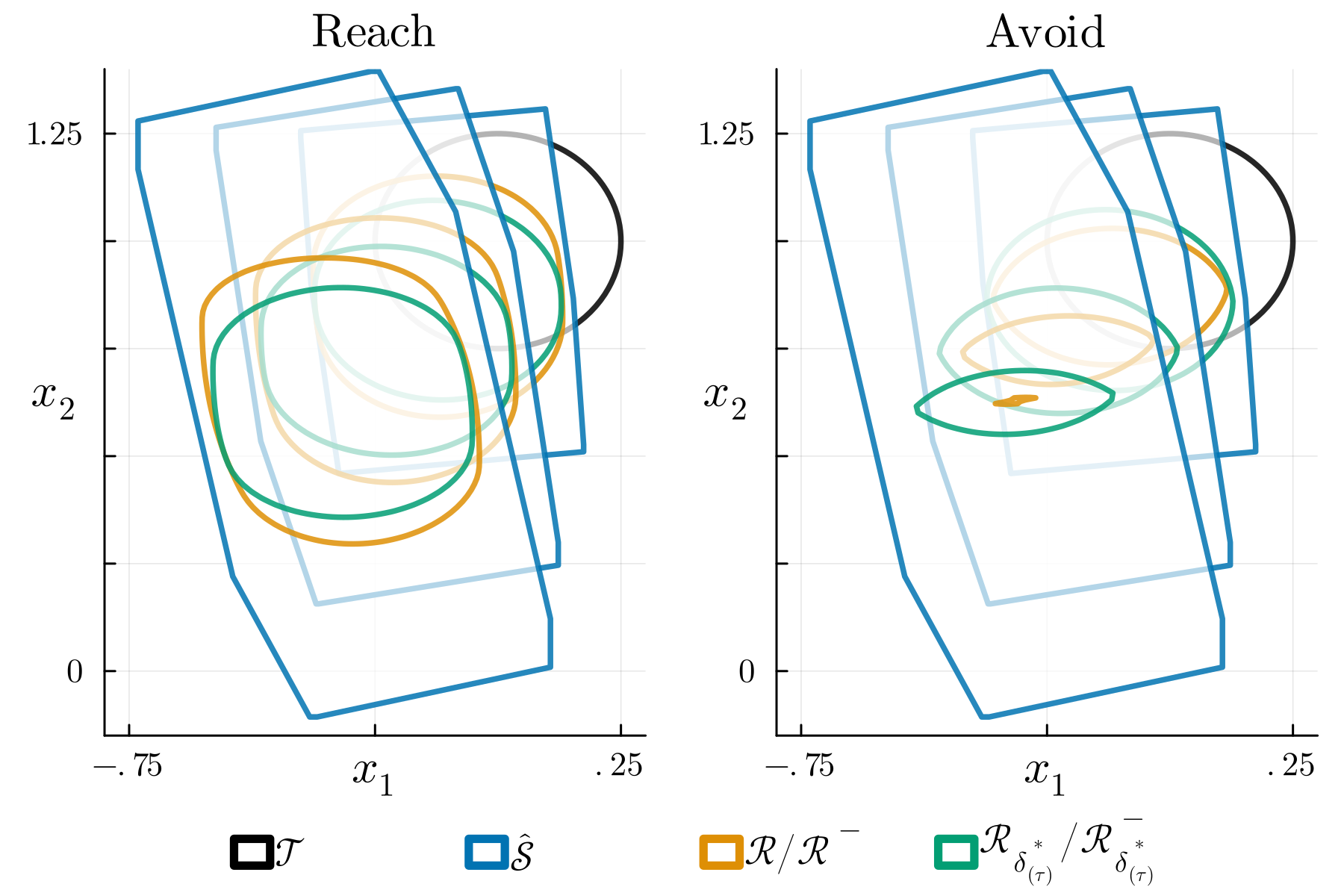}    
    \caption{\label{fig: front_fig} \textbf{Time-Varying Safe Linear Envelopes for a Differential Game in the Van der Pol System \eqref{vdpdynamics}.} For a target $\tset$ (black), true reachable sets $\hjrset/\hjrset^-$ (gold) are solved for a \textit{Reach} and \textit{Avoid} game respectively at $t = [0.13, 0.26, 0.39]$. The corresponding reachable sets of the safe envelope $\hjrset_{\maxerr(\tau)}/\hjrset^-_{\maxerr(\tau)}$ (pink) based on the time-varying Taylor-series linearization error. Compared with Fig.~\ref{fig:basic}, the safe envelope with the time-varying error is tighter than that of the constant error as guaranteed by Cor.~\ref{cor:tvFS}. Temporal slices of the conservative approximation $\fseto$ of the feasible tube $\ftube$ \cite{chen2012taylor, JuliaReach19} are shown at times $t\in [-0.13, -0.26, -0.39]$ (blue). }
    \label{fig:TVonly}\vspace{-1em}
\end{figure}

To use the state-independent Hopf formula, the error $\maxerr$ cannot be a function of the state $x$, however, we may define a time-varying error bound $\maxerr(\tau):[t,T]\mapsto\mathbb{R}$ that implicitly captures a spatial relationship for a tighter bound on the error. The following result is demonstrated in Fig.~\ref{fig:TVonly}).

\begin{Corollary} (Time-varying Error) \label{cor:tvFS} \\ 
Let the time-varying maximum error be defined by,
\begin{eqnarray}
\maxerr(\tau) \triangleq \max_{\ftube(\tset, \tau) \times \Sigma (\tau)} \big \Vert [\sys - \lsys](x,u,d, \tau) \big \Vert, \tau \in [t,T].
\end{eqnarray}
Then for \textit{Reach} and \textit{Avoid} with $\ham_{\maxerr(\tau)}$ and $\ham_{\maxerr(\tau)}^-$ respectively,
\begin{eqnarray}
\begin{aligned}
\hjrset_{\maxerr}(\tset,\tau) &\subset \hjrset_{\maxerr(\tau)}(\tset,\tau) \subset \hjrset(\tset,\tau), \\
\hjrset_{\maxerr}^-(\tset,\tau) &\supset \hjrset_{\maxerr(\tau)}^-(\tset,\tau) \supset \hjrset(\tset,\tau).
\end{aligned}
\end{eqnarray}
If $\ftube(\tset, \tau)$ is evaluated sparsely at $\{s_i\}_{s_1=t}^{s_n=T}$ then the function $\hat\maxerr(\tau)\triangleq\{ \maxerr(s_i^*) \:|\: s_i^* \triangleq \max_{s_i \le \tau} s_i\}$ will have the same intermediate quality.
\end{Corollary}

\begin{proof}
By Thm.~\ref{thm:errenv}, the games defined with $\maxerr(\tau)$ must satisfy the RHS $\forall \tau\in [t,T]$. For the LHS, the set of all trajectories monotonically expands for increased time-intervals \cite{filippov1960differential}, hence $\forall \tau \in [t,T], \ftube(\tset, \tau) \subseteq \ftube(\tset, t)\implies \maxerr(\tau) \le \maxerr$. This argument also holds for discrete intervals $[s_{i}, s_{i+1}]$, thus, $\maxerr(\tau) \le \hat \maxerr(\tau) \le \maxerr$.
\end{proof}


\subsection{Determination of $\fset$ and $\maxerr$} \label{sec:det}

The guarantees throughout this paper hinge on the ability to define the feasible tube $\ftube$ and corresponding error $\maxerr$ between the linearized and true system over the domain $\ftube$. While the existence of $\ftube$ and $\maxerr$ may be known, computing a guaranteed superset $\fseto \supset \ftube$ and then solving (\ref{maxerrdeftube}) are separate feats. For the conservative approximation of $\ftube$, there are several works \cite{althoff2008reachability, althoff2011zonotope, chen2012taylor} outlining efficient methods that might be employed, generally by conservatively expanding a convex polytope covering $\fset$ over a discrete time interval. For the computation of $\maxerr$, a sampling-based method might be derived, however it would transform all previous guarantees into probabilistic guarantees, and we leave this for future work. Some linear models may require system-specific derivations of the error or optimization procedures, however, for the Taylor series linearization, a general derivation serves as a foundation and is given in the Supplementary (see \href{https://drive.google.com/file/d/13gXlDd5__ZoE15QeyZeEQm3pa_vlUxzI/view?usp=sharing}{online}).

\section{CONTROL OF A MULTI-AGENT DUBIN'S PURSUIT-EVASION GAME}\label{sec:Demos}

Consider a system of $N$ Dubin's cars in the relative frame of an additional car as posed in \cite{mitchell2005time, Kirchner_2018}, such that each agent's relative state, $x_{i} = [x_{\Delta_i}, y_{\Delta_i}, \theta_{\Delta_i}]$, evolves by
\begin{eqnarray}
    \begin{aligned}
        \dot x_{i} = \begin{bmatrix}
        -v_a + v_b \cos(\theta_{\Delta_i}) + a {y_{\Delta_i}} \\
        v_b \sin({\theta_{\Delta_i}}) - a {x_{\Delta_i}} \\
b_i - a
        \end{bmatrix}
    \end{aligned}
\end{eqnarray}
where $a$ and $b_i$ are the controls of an evading agent and $i \in [1, N]$ pursuing agent respectively, $v_a$ and $v_b$ are their velocities respectively, and $\theta_i$ is the relative heading of the $i \in [1, N]$ agent. Let the game be defined s.t. the evader seeks to avoid all $i \in [1,N]$ pursuers while each pursuer aims to reach the evader. The action of the evader couples the dynamics, barring the application of decomposition methods. 

The linearization of this system takes the form \cite{Kirchner_2018}
\begin{eqnarray}
    \begin{aligned} \dot x= & {\left[\begin{array}{ccc}A_1(s) & \cdots & 0 \\ \vdots & \ddots & \vdots \\0 & \cdots & A_N(s) \end{array} \right]\left[\begin{array}{c}x_1 \\x_2 \\\vdots \\x_N\end{array}\right] } 
    +\left[\begin{array}{c}\varepsilon_1 \\\varepsilon_2 \\\vdots \\\varepsilon_N\end{array}\right] \\& 
    +\left[\begin{array}{c}B_{a_1}(s) \\B_{a_2}(s) \\\vdots \\B_{a_N}(s) \end{array}\right] a+\left[\begin{array}{ccc}B_{b_1} & \cdots & 0 \\ \vdots & \ddots & \vdots \\0 & \cdots & B_{b_N}\end{array}\right]\left[\begin{array}{c} b_1 \\b_2 \\\vdots \\b_N \end{array}\right]
    , \end{aligned}
    \label{dubins_linearized}
\end{eqnarray}
where $(A_i(s), B_{a_i}(s), B_{b_i}) \triangleq (D_x f, D_a f, D_b f) |_{\tilde {x}_i, 0, 0}$ and $\varepsilon_i$ is the residual error. The error bound is derived via the method proposed in Sec.\ref{subsec:TSlin} by linearizing around the evader trajectory with no action, making use of Cors. ~\ref{cor:tvFS}, ~\ref{cor:spFS} and ~\ref{lem:FBerr} to tightly bound the value.

Let the team of $N$ pursuers surround the evader initially in the form of a regular $N$-polygon, heading toward the evader. The initial position and angle of each agent is then given by 
\begin{eqnarray}
\begin{aligned}
    \begin{bmatrix} x_{\Delta_i,0} \\ y_{\Delta_i, 0} \end{bmatrix} &= R_{\text{rot}}\left((i-1)\frac{2 \pi}{N} + \theta_a \right) \begin{bmatrix} 0 \\ r_p \end{bmatrix}
    \\ \theta_{\Delta_i, 0} &= (i-1)\frac{2 \pi}{N} + \theta_a - \pi/2 
\end{aligned} \label{ic}
\end{eqnarray}
where $\theta_a$ is the angle of the evader and $r_p$ is the initial distance between each pursuer and the evader and is defined s.t. $ r_p = v_a = v_b = 3$.


\begin{figure}[t]
    \centering
    \includegraphics[width=\linewidth]{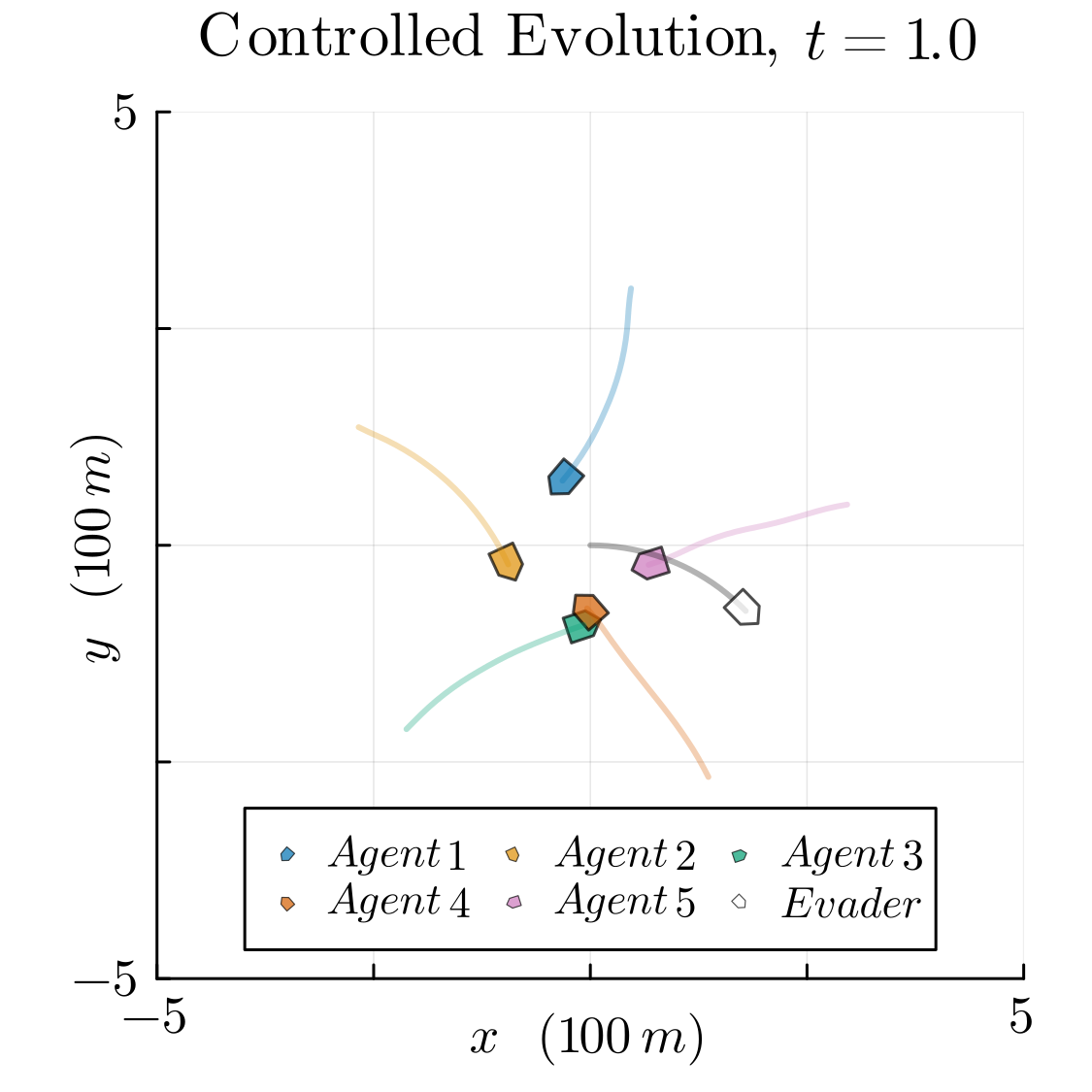} 
    \caption{\textbf{Evasive Evolution in the $5$ agent (15D) Multi-Agent Dubin's Systems with the Hopf Controller.} Despite capture being labeled as feasible (i.e. there exists a control and disturbance that \textit{can} lead to capture) for the initial condition, the value from the Hopf solution $V_{\maxerr}^-(x,s) > 0, \forall s \in [0,1]$, implying that capture is avoidable under the \textit{optimal} control from the evader. Namely, capture may be avoided with the policy computed from $\nabla V_{\maxerr}^-$ in (\ref{HJoc}), which takes $0.15s$ to solve entirely and $<1e-3s$ to evaluate online.}
    \label{fig:MD5_sim}\vspace{-1em}
\end{figure}

In this demonstration, we assume the perspective of the evader with the goal of  computing the evasive maneuver to avoid capture. We assume that capture may occur if and only if the evader is in close proximity ($r\le1/2$) to \textit{any} pursuer. Hence, the target for each pursuer takes the form of an ellipsoid, given by
\begin{eqnarray}
    \Omega_i \triangleq \{ x\: |\: x^\top W^{-1}_i x \le 1 \},
\end{eqnarray}
where $W_i \in \mathbb{R}^{3N \times 3N}$ is a block-diagonal matrix of diagonal matrices with values of $r^2$ in the $i$-th block's first and second diagonals and $\theta_{max}^2$ in the third, and $r_{max}^2$ in the $j$-th block's first and second diagonals $\theta_{max}^2$ in the third for all $j\neq i$. The joint system has a target then defined by $\Omega = \cup \Omega_i$ and the resulting target function $J$ leads to a special $V^-$ \cite{Kirchner_2018}:
\begin{eqnarray} 
    J(x, t) = \min_{i \in [1, N]} J_i(x) \implies  V^- (x, t) = \min_{i \in [1, N]} V_i^- (x, t),
\end{eqnarray}
which max-plus algebras certify may be solved independently (and in parallel) \cite{darbon2016algorithms}. The optimal controls for the team will then be determined from the minimum value, corresponding to whichever pursuer is most able to capture from the initial condition $(x,t)$. We assume the evader must escape capture by any agent for up to $t=1s$ (beyond which capture is infeasible due to the velocities), and hence seek to certify
\begin{eqnarray}
    \min_{s \in [0, 1]} V^- (x, s) > \min_{s \in [0, 1]} V_{\maxerr}^- (x, s) > 0
\end{eqnarray}
where $V_{\maxerr}^-$ is the value of the game with the linearized dynamics and antagonistic error proposed in Sec.~\ref{sec:SfEnv}. The Hamiltonian may be non-convex, however, due to the nature of the \textit{Avoid} game, a safe lower-value $V^-_{\maxerr}$ is always given by the minimax-viscosity solution (Rem.~\ref{lem:cvxyH}). 



\begin{figure}[b!]
    \centering
    \includegraphics[width=\linewidth]{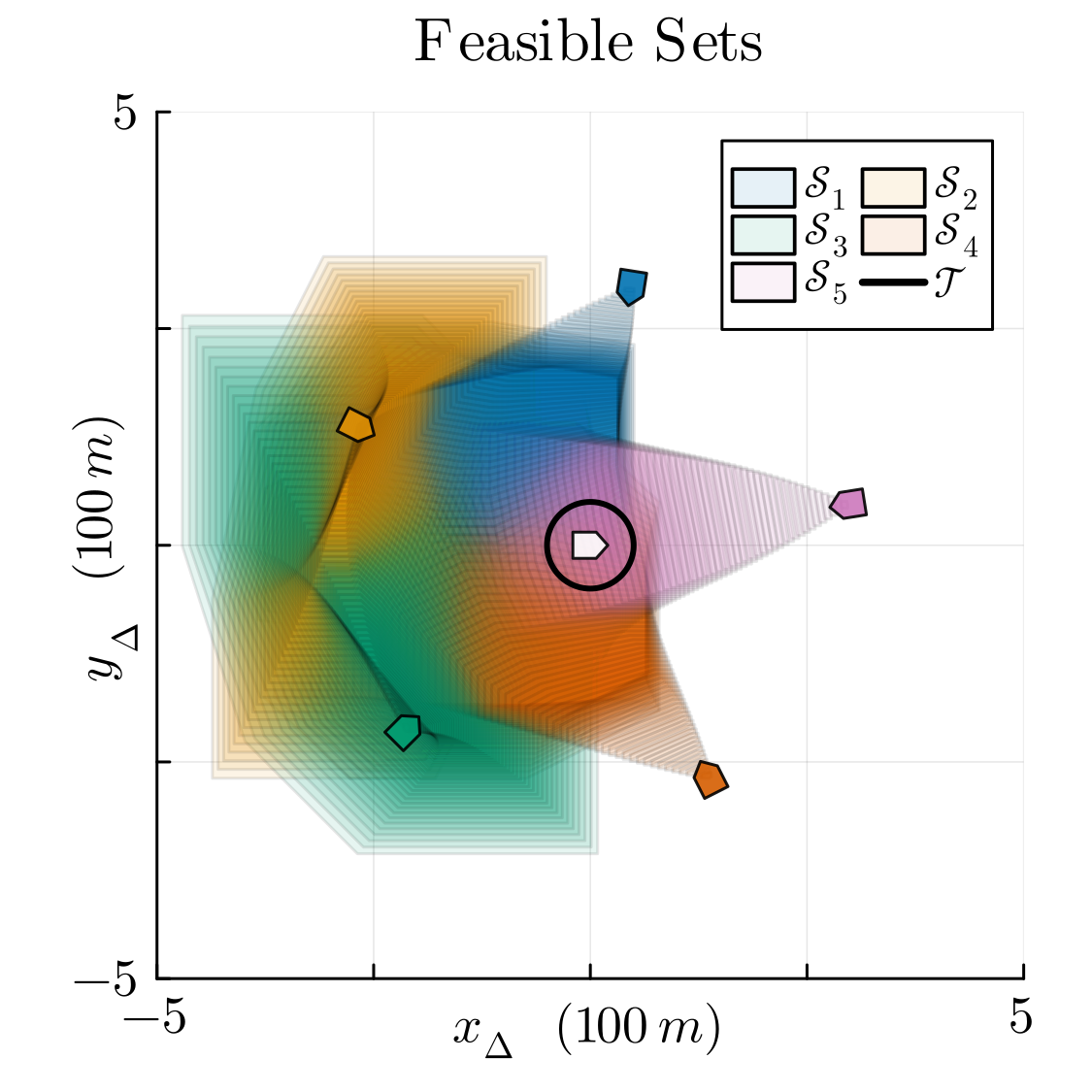} 
    \caption{\textbf{Pursuit-Evasion Feasibility in the 5 agent (15D) Multi-Agent Dubins Systems} The pursuers seek to enter the target set (black circle) to capture the evader. The projection of temporal slices from $t\in[0, 1]$ of the conservative sets $\fseto_i \supset \ftube_i$ (computed with TMs \cite{chen2012taylor, JuliaReach19}) for each agent are shown in the relative states $(x_\Delta ,y_\Delta)$. The union of pursuer feasible sets contains the evader starting from $t=0.3$; the feasibility benchmark alone cannot guarantee escape.}
    \label{fig:MD5_feas}\vspace{-1em}
\end{figure}

For these reasons, we use the Hopf formula with the linearization envelope error to certify where capture is avoidable for an $N=5$ system. Conservative feasible sets of each pursuer $\fset_i$, i.e. temporal slices of the over-approximated tube $\fseto_i \supseteq \ftube_i$, are computed with the TM zonotope method \cite{chen2012taylor} and used to generate the discrete, time-varying antagonistic error (see Cor.~\ref{cor:tvFS}) for the initial condition $\theta_a=-0.16$. The game (15D) is too high-dimensional to solve with standard DP-based methods, however, we benchmark the performance against the TM zonotope method \cite{chen2012taylor}, which outperforms all popular DI methods at the time of writing \cite{JuliaReach19}. The sets of the benchmark are shown in the Supplementary (see \href{https://drive.google.com/file/d/13gXlDd5__ZoE15QeyZeEQm3pa_vlUxzI/view?usp=sharing}{online}, Fig.~\ref{fig:MD5_feas}).

In this configuration, the benchmark dictates that capture is \textit{feasible} for several agents, i.e. there exists a set of control inputs for both the evader and pursuers that lead to capture. However, the Hopf-based solution at the initial position in \eqref{ic} gives a value greater than zero, indicating that evasion is possible.
Moreover, only the latter directly synthesizes a control policy paired with the solution. This example underscores the benefit of the differential game-based method proposed in this work in comparison with fast feasibility methods. While losing trajectories are often possible, taking advantage of the game nature offers improved conservative linearization approaches.

To verify these results, we simulate the safe evasive policy yielded by the Hopf method against pursuers driven by a model predictive controller \cite{bemporad2002model} which uses the same time-varying linearization in \eqref{dubins_linearized} but samples the error $\varepsilon_i \in \errset$. We chose this opponent to demonstrate an independent baseline; the Hopf-based pursuer policy is guaranteed to fail by Thm.~\ref{thm:errenv}. The trajectory is plotted in Fig.~\ref{fig:MD5_sim}. The Hopf formula takes $0.15s$ to solve the value and compute the entire evasive policy and evaluates in less than $1\mathrm{e}{-3}s$ online.

\section{Conclusion}\label{sec:conclusion}

We have demonstrated that the linear error for a nonlinear system may be transformed to give guarantees on high-dimensional, nonlinear optimal control and differential games.
Ultimately, the proposed work offers a novel method for solving high-dimensional HJR with guarantees, which proves to be less conservative than feasibility methods alone. 
Future work includes extensions to high-dimensional lifted spaces and relaxation to probabilistic guarantees. 



\section*{Acknowledgment}

We thank Somil Bansal and Donggun Lee for discussions. We thank Dylan Hirsch, Zheng Gong, Nikhil Shinde and Sander Tonkens for valuable feedback on the paper.


\section*{References}
\vspace{-1em}

\bibliographystyle{IEEEtran}
\bibliography{main}

\clearpage 

\section*{Supplementary}
\label{sec:apx}

\section{Practical Considerata} \label{subsec:lim}

As described in Sec.~\ref{subsec:hopf}, the Hopf formula given in (\ref{HopfFormula}) yields the viscosity solution when $J$ and $\ham$ are convex. However, the addition of the antagonistic error may corrupt this quality, depending on the relative sizes of $\cset$, $\dset$, and $\errset$, i.e. based on the size of the control authority, which is generally a fixed quality of a system.

\begin{Remark} (Convexity of the $\maxerr$-Hamiltonians) \label{lem:cvxyH} \\
Assume the linear Hamiltonian's $\haml, \haml^-$ are convex (this may be checked for any system by the conditions in \cite{sharpless2023koopman}). These are transformed by the addition of error in the following way,
    \begin{eqnarray}
        \begin{aligned}
            \ham_{l, \mathcal{Z}} &\triangleq \mathcal{I}^\star_\cset (R_1(t)p) -\mathcal{I}^\star_\dset (R_2(t)p) \\
            &\Rightarrow \ham_{\maxerr, \mathcal{Z}} = \ham_{l, \mathcal{Z}} -\mathcal{I}^\star_\errset (- \Phi(T-\underline{t})^{-\top} p), \\
            \ham_{l, \mathcal{Z}}^- &\triangleq \mathcal{I}^\star_\dset(R_2(t)p) -\mathcal{I}^\star_\cset(R_2(t)p) \\
            &\Rightarrow \ham_{\maxerr, \mathcal{Z}}^- = \ham_{l, \mathcal{Z}}^- +\mathcal{I}^\star_\errset (- \Phi(T-\underline{t})^{-\top} p). \\
        \end{aligned}
    \end{eqnarray}
    Therefore,
    \begin{enumerate}
       \item In the two-player \textit{Reach} game, $\ham_{\maxerr, \mathcal{Z}}$ will not necessarily be convex and must be checked for a specific definition of $\cset, \dset$ and observed error $\errset$.
       \item In the two-player Avoid game, $\ham_{\maxerr, \mathcal{Z}}^-$ will be convex for any $\maxerr \ge 0$.
    \end{enumerate} 
    \label{thm:HamCvxyCond}
\end{Remark}

The conditions in Remark~\ref{thm:HamCvxyCond} have significant implications on any applications of Thm.~\ref{thm:errenv} with the Hopf formula and the methods proposed in the work, particularly to \textit{Reach} games. In \textit{Avoid} games, the error does not alter convexity and thus the proposed work may be used for safe controller synthesis in any \textit{Avoid} problem. 



In addition to the limitations of the Hopf formula, one must note a large maximum error will also yield poor guarantees. While the solution is guaranteed to be optimal with respect to the true dynamics, it is also optimal with respect to any $\maxerr$-\textit{similar} dynamics. The optimal trajectory of the new game will take the path which is most robust to disturbance \textit{and error}. To ameliorate this burden, we consider a few methods for decreasing the distance between the envelope boundary and the true set.




\section{Methods for Tighter Safe Envelopes} \label{sec:improvements}

We propose a few methods for reducing the conservativeness of our method. 
For clarity, each is described independently, however, as demonstrated in the running example, none is incompatible with another.

\subsection{Disturbance-only Feasibility and Error} \label{subsec:sperr}

\begin{figure}[t]
    \centering
    \includegraphics[width=\linewidth]{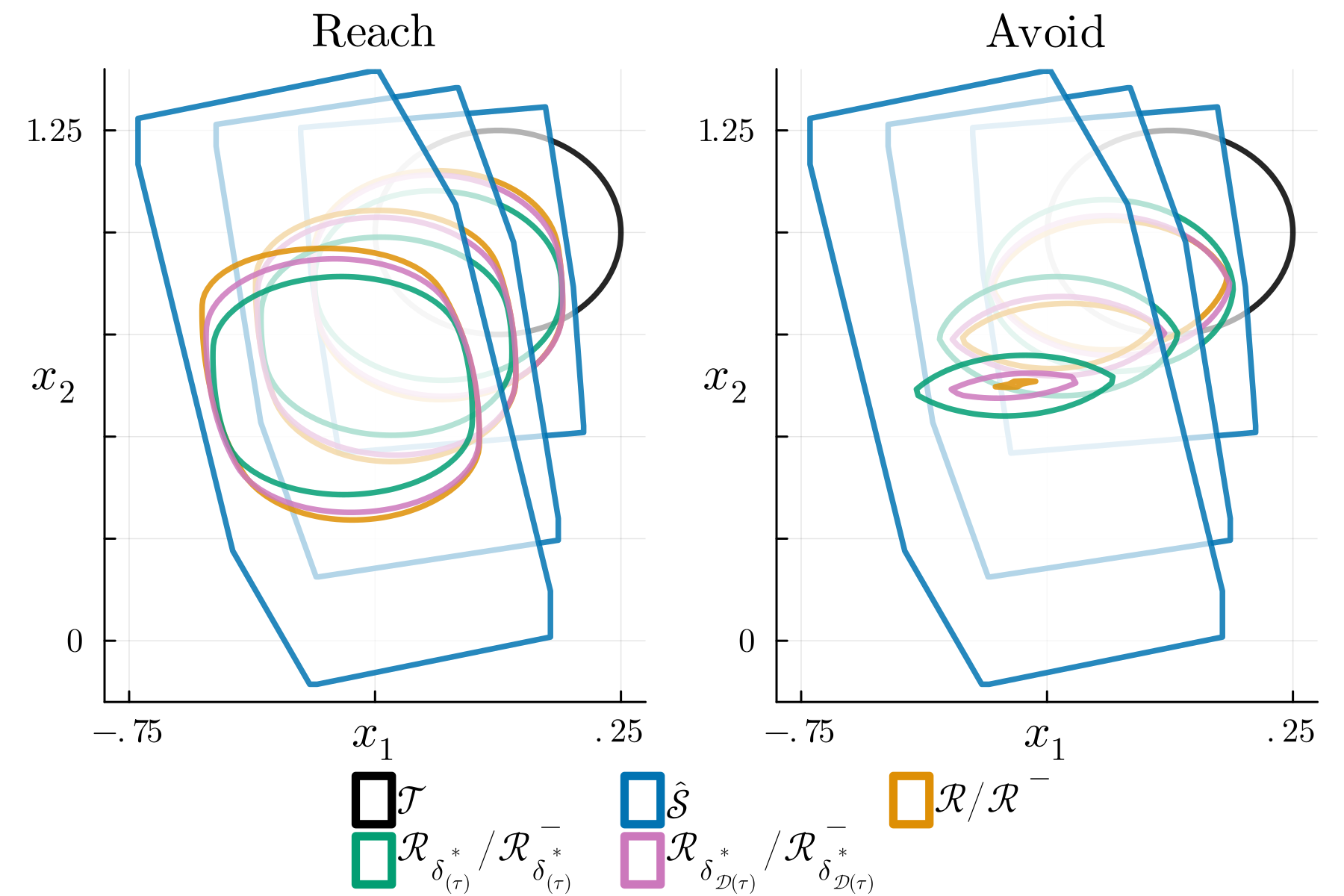}    
    \caption{\textbf{Disturbance-Only Safe Envelopes in the Van der Pol System for a Two-Player Game} For a target $\tset$ (black), true reachable sets $\hjrset/\hjrset^-$ (gold) are solved for a \textit{Reach} and \textit{Avoid} game respectively at $t = [0.13, 0.26, 0.39]$. The corresponding reachable sets of the safe envelope $\hjrset_{\maxerr_\dset(\tau)}/\hjrset^-_{\maxerr_\dset(\tau)}$ (pink) based on the time-varying Taylor-series linearization error for \textit{disturbance-only} feasible sets are also solved. The reachable sets of the time-varying safe envelope $\hjrset_{\maxerr(\tau)}/\hjrset^-_{\maxerr(\tau)}$ (green) from Fig.~\ref{fig: front_fig} are included for comparison. Cors. \ref{cor:tvFS} and \ref{cor:spFS} guarantee that each of the envelopes (pink and green) will be tighter than the basic constant-error envelope (Fig~\ref{fig:basic}) and we may see combining them further improves this quality. The corresponding temporal slices of the conservative approximation $\fseto \supset \ftube$ \cite{chen2012taylor, JuliaReach19} are included (blue).}
    \label{fig:TVSIS}\vspace{-1em}
\end{figure}

Recall that the feasible tube $\ftube$ to bound the error is defined as the set from which there exist some inputs in $\cset$ and $\dset$ such that the system may arrive at the target $\tset$. In other words, it allows the control and disturbance to work together to reach the target set, leading to a potentially very large over-approximation of the true reachable sets. One may note from the proofs of Thm.~\ref{thm:errenv} and Cor.~\ref{cor:saferr}, to assume the set $\ftube$ contains all trajectories evolving to $\tset$ for $\csetsig \times \dsetsig \times [t,T]$ is sufficient but superfluous. The proof illustrates the error need only be defined on the trajectories evolving from an $\sigma$-near optimal disturbance strategy. While this set cannot be localized without the solution, we show it is contained within the set for the game without the control, since control (its opponent) only reduces this set.
This is valuable because the disturbance-only tube $\ftube_\dset$ is a subset of $\ftube$ for both players (and often significantly smaller), and hence reduces the antagonistic error. 
Note, the lack of control guarantees a tighter feasible set that yields a conservative antagonistic error, but the game still includes control.
This is formalized in the following and demonstrated in the Van der Pol system in Fig.~\ref{fig:TVSIS}.

\begin{Corollary} (Disturbance-only Error) \label{cor:spFS} \\
Suppose $\ftube_\dset$ is defined as in (\ref{rsetdef}) for the disturbance-only ($\cset=\{0\}$) setting, and let the maximum error on this set $\maxerr_\dset$ be defined as \eqref{maxerrdeftube}. Then for \textit{Reach} and \textit{Avoid} games respectively,
\begin{eqnarray}
\begin{aligned}
\hjrset_{\maxerr}(\tset,\tau) &\subset \hjrset_{\maxerr_\dset}(\tset,\tau) \subset \hjrset(\tset,\tau), \\
\hjrset_{\maxerr}^-(\tset,\tau) &\supset \hjrset_{\maxerr_\dset}^-(\tset,\tau) \supset \hjrset(\tset,\tau).
\end{aligned}
\end{eqnarray}
\end{Corollary}
\begin{proof}
We assume $0 \in \cset$ w.l.o.g. as we may always define $\hat f \triangleq f + c$ with $u \in \hat\cset\triangleq(\cset - c)$ for $c \in \cset$. Since the trivial input is an element of the input set, $\fset_\dset \subset \fset $ and thus $\maxerr_\dset \le \maxerr$, proving the LHS above. With $\ftube_\dset$, we may apply Cor.~\ref{cor:saferr} to either the \textit{Reach} or \textit{Avoid} games to know,
$$
\hjrset_{\maxerr_\dset}(\tset,\tau) \subset \hjrset_\dset(\tset, \tau), \quad \& \quad \hjrset_{\maxerr_\dset}^-(\tset,\tau) \supset  \hjrset_\dset^-(\tset, \tau).
$$
But of course, disturbance itself creates an envelope (assuming w.l.o.g. $0 \in \dset$) i.e., 
$$
\hjrset_\dset(\tset, \tau) \subset \hjrset(\tset, \tau), \quad \& \quad \hjrset_\dset^-(\tset, \tau) \supset  \hjrset^-(\tset, \tau),
$$
and, hence the RHS is proved. 
\end{proof}

\begin{figure*}[t]
    \centering
    \includegraphics[width=\linewidth]{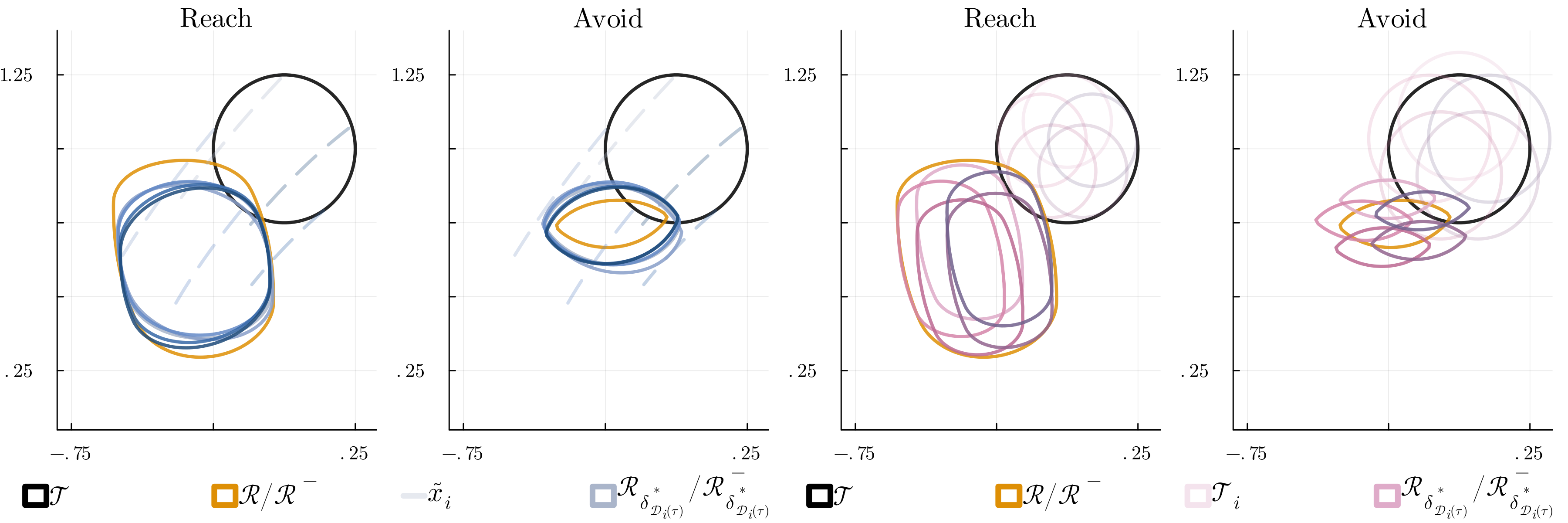}    
    \caption{\textbf{Linear Ensemble and Target Partitioning Safe Envelopes in the VanderPol System} For a target $\tset$ (black), the true reachable sets $\hjrset/\hjrset^-$ (gold) are solved for a \textit{Reach} and \textit{Avoid} game at $t = 0.26$ \& $ t=0.39$ respectively using dynamic programming. Left plots: the corresponding reachable sets of the linear ensemble envelopes with maximum error $\maxerr_{\dset_i}$ (blues) based on the ensemble trajectories (blue dashed), with time-varying, disturbance-only Taylor-series linearization error are solved. Cor. ~\ref{lem:ens} guarantees that the union (for the \textit{Reach} game) and intersection (for the \textit{Avoid} game) of ensemble envelopes are also safe envelopes. Right plots: the corresponding reachable sets of the partition envelopes with $\maxerr_{\dset_i}$ (pinks) based on the partitioned target $\tset_i$ (faint pinks), with time-varying, disturbance-only Taylor-series linearization error are solved. Cor. ~\ref{lem:par} guarantees that the union of partition envelopes in \textit{Reach} and \textit{Avoid} games is also a safe envelope. 
    }
    \label{fig:LETP}\vspace{-1em}
\end{figure*}

\subsection{Ensemble Errors} \label{subsec:leerr}

We may also take advantage of the fact that variations of the problem at hand can yield unique safe envelopes with respect to the original game and hence their results can be consolidated to generate a tighter bound of the true set. One simple variation to consider is of the linear model, perhaps due to a different linearization parameter or method. Each linearized model (with appropriately defined error) will lead to a different conservative envelope of the true reachable set. Therefore these sets can be combined to reduce conservativeness.

\begin{Corollary} (Ensemble of Linear Models) \\ \label{lem:ens}
Let $\{\lsys_i(x,u,d, \tau)=A(\tau)_i x + B_{1,i}(\tau)u + B_{2,i}(\tau)d + c_i\}_i$ be a set of linear models and $\maxerr_i$ be the maximum error for model $\lsys_i$ on $\ftube$. Then for reach and avoid games,
\begin{eqnarray}
\bigcup_i \hjrset_{\maxerr_i}(\tset,\tau) \subset \hjrset(\tset,\tau) \subset \bigcap_i \hjrset_{\maxerr_i}^-(\tset,\tau).
\end{eqnarray}
\end{Corollary}
\begin{proof}
By Thm.~\ref{thm:errenv} and Cor.~\ref{cor:saferr}, $x \in \hjrset \supset \hjrset_{\maxerr_i} \implies x \in \bigcup_i \hjrset_{\maxerr_i}$, and if $x \in \hjrset \subset \hjrset_{\maxerr_i}^- \implies x \in \bigcap_i \hjrset_{\maxerr_i}^- $. 
\end{proof}

Notably, the sub-problems corresponding to each linear model may be solved independently, and hence the load of the total ensemble solution scales in a parallelizeable, linear fashion with respect to number of models. This result is demonstrated in the Van der Pol system in Fig.~\ref{fig:LETP}.

\subsection{Forward Feasibility and Error} \label{subsec:locerr}

There may be a substantial reduction in the antagonistic error in the special case when considering initial conditions in a local region given by a bounded set $\localX \subset \ssp$, e.g. a single state. Note, this is a frequent case which occurs when the initial condition is fixed, such as in feedback control online and forecasting problems. We may take advantage of the fact that any trajectory from the current point to a target must be within not only the backward feasible tube but also the forward feasible tube.

Let the \textit{forward} feasible set $\fset^+$ of a set $\localX$ at time $T$ ahead of the current time $t$, be given by
\begin{eqnarray}
\begin{aligned}
\fset^+(\localX, T) \triangleq \{x \:\:  |& \:\:  \exists u(\cdot)\in \csetsig \:\: \exists d(\cdot)\in \dsetsig \\ &\text{ s.t. } \tj(t) \in \localX, \tj(T) = x \}.
\label{frsetdef}
\end{aligned}
\end{eqnarray}
Moreover, the forward feasible tube $\ftube^+$ is then given by
\begin{eqnarray}
\begin{aligned}
\ftube^+(\localX, s) \triangleq \bigcup_{\tau \in \tint} \fset^+(\localX, \tau).
\label{fftubedef}
\end{aligned}
\end{eqnarray}
These are analogous to \eqref{rsetdef} \& \eqref{ftubedef} respectively but are more relevant in the following corollaries.

\begin{Corollary}
(Forward Error) \label{lem:FBerr} \\Let $\ftube^+(\localX, T)$ be the future feasible tube of $\localX$ at $T$ time in the future and $\maxerr_g$ be the error. Then for \textit{Reach} and \textit{Avoid} games respectively,
\begin{eqnarray}
\begin{aligned}
&V_{\maxerr_g} (x,t) \le \bar c \implies V (x,t) \le \bar c, \\
&V_{\maxerr_G}^- (x,t) \ge \bar c \implies V^- (x,t) \ge \bar c.
\end{aligned} \quad x \in \localX
\end{eqnarray}
\end{Corollary}
\begin{proof}
This is a simple application of Thm.~\ref{thm:errenv}, if $\exists \tj$ s.t. $\tj(t)=x$ \& $\tj(T) \in \tset$, then $\tj(\tau) \in \ftube^+(\localX, T)$.
\end{proof}


\subsection{Partition Errors} \label{subsec:tperr}

Lastly, another significant variation one may make for reduced conservativeness is of the target set. Recall, the error bound is defined based on the feasible tube $\ftube$ containing all trajectories that may enter the target set $\tset$. If $\tset$ is covered by a set, say a \textit{partition}, of smaller targets, say \textit{elements}, one may consider the feasible tube for each element. The maximum error over each of the smaller tubes will be less than or equal to the maximum error over the original tube, thereby freeing some divisions from the maximum error in another. 
For the \textit{Avoid} game, an assumption on the distribution of the elements is necessary.


\begin{Corollary}
(Target Partitioning) \\ \label{lem:par}
Suppose $\{\tset_i\}$ is a partition of $\tset$ such that $\cup_i \tset_i \subseteq \tset$ and $\tset_i \subseteq B(\epsilon), \epsilon>0$, a ball of radius $\epsilon$. Let $\bar\fset_i$ be defined as in (\ref{rsetdef}) with $\tj(T)\in \tset_i$, and let $\maxerr_i$ be the corresponding error for $\lsys$. Then for the Reach game,
\begin{eqnarray}
\begin{aligned}
\hjrset_{\maxerr}(\tset,\tau) &\subset \bigcup_i \hjrset_{\maxerr_i}(\tset_i,\tau) \subset \hjrset(\tset,\tau). \\
\end{aligned}
\end{eqnarray}
For the Avoid game, suppose $\{\tset_i\}$ is a partition of  $\tset$ such that $\cup_i \tset_i \supset \tset$ and $\tset \supseteq B(\epsilon), \epsilon>0$. 
Then if for any $x \in \fset(\tset_i, t) \cap \fset(\tset_j, t)$, it happens that the future feasible set of $x$ at time $T$ satisfies $\fset^+(x, T) \subseteq \tset_k$ for $\tset_k \in \{\tset_i\}$, the following property holds,
\begin{eqnarray}
\begin{aligned}
\bigcup_i \hjrset_{\maxerr_i}^-(\tset_i,t) \supset \hjrset^- (\tset,t). \\
\end{aligned}
\end{eqnarray}
\end{Corollary}
\begin{proof}
For the \textit{Reach} game, by definition $\tset_i \subseteq \tset$ so $\ftube_i \subseteq \ftube \implies \maxerr_i \le \maxerr$. Then Thm.~\ref{thm:errenv} and Cor.~\ref{cor:saferr} may be applied to conclude that 
\begin{eqnarray}
\begin{aligned}
\hjrset_{\maxerr}(\tset_i,\tau) \subset \hjrset_{\maxerr_i}(\tset_i,\tau) \subset \hjrset(\tset_i,\tau). \\ 
\end{aligned}
\end{eqnarray}
Since $\cup_i \tset_i \subseteq \tset$ then for any $\delta\in[0, \delta^*], \cup_i \hjrset_\delta(\tset_i,\tau) \subseteq \hjrset_\delta(\tset,\tau)$ because $x \in \cup_i \hjrset_{\delta_i}(\tset_i,\tau) \implies \exists i \text{ s.t. } \tj(T) \in \tset_i \implies \tj(T) \in \tset \implies x \in \hjrset_\delta(\tset,\tau)$.
Thus, we may take the union over $i$ across the relations above and arrive at the result.

In the \textit{Avoid} setting, we must be more cautious. While Thm.~\ref{thm:errenv} and Cor.~\ref{cor:saferr} may be applied to conclude that $\hjrset^-_{\maxerr}(\tset_i, \tau) \supset \hjrset^-(\tset_i, \tau)$, in general $\cup_i \hjrset^-(\tset_i, \tau) \not \supset \hjrset^-(\tset, \tau)$ since $\tj(\tau; x)$ may ``dodge'' partition elements independently but still be within their union. In this case, the given assumption is a sufficient condition. For proof, let $\exists y \in \hjrset^-(\tset, t)$ but $\exists y \notin \hjrset^-(\tset_i, t)$. But $\cup_i \tset_i \supset \tset$ hence, $y \in \fset(\tset_i, t) \cap \fset(\tset_j, t)$. It follows by assumption then $\fset^+(y, t) \subset \tset_k$ which implies $y \in \hjrset^-(\tset_k, t)$ and thus gives a contradiction. \qedhere

\end{proof}


Note, to guarantee that a target is unavoidable (in the \textit{Avoid} game), the entire partition solution needs to be checked, however, to check if a target is reachable (in the \textit{Reach} game) requires only one partition to be reachable. 
Ultimately, the division of space and the resulting increase in local accuracy greatly improves the total union guarantee, giving the closest experimental results to the true solution. This is demonstrated in the running example in Van der Pol system in Fig.~\ref{fig:LETP}.

\subsection{Taylor Series Linearization for Bounds on $\maxerr$}\label{subsec:TSlin}

Consider the Taylor series of $\sys$ in (\ref{Dynamics}) about a trajectory $\tilde \tj(\tau)$ defined by $\tilde \tj(t) = \tilde x$, $\tilde u(\cdot) \in \csetsig$, $\tilde d(\cdot) \in \dsetsig$,
\begin{eqnarray}
\begin{aligned}
\dot{x} = \sys(p) + &A(\tau)(x - \tilde \tj(\tau)) + B_1(\tau)(u - \tilde u(\tau))\\ + & B_2(\tau)(d - \tilde d(\tau)) + \eps_{\tilde \tj}(x),
\end{aligned}
\end{eqnarray}
for some error $\eps_{\tilde \tj}$ and 
$$A(\tau), B_1(\tau), B_2(\tau) \triangleq  D_xf|_{\tilde \tj(\tau)}, D_uf|_{\tilde \tj(\tau)}, D_df|_{\tilde \tj(\tau)}.$$ Let $x_\Delta (\tau) \triangleq x - \tilde \tj(\tau)$ (and same for $u_\Delta$ \& $d _\Delta$) define the shifted space. Note, the Taylor series for $\dot{x}_\Delta$ around its origin implies  $(x, u, d)=(\tilde \tj, \tilde u, \tilde d)$ and takes the form,
\begin{eqnarray}
\dot{x}_\Delta = A(\tau)x_\Delta + B_1(\tau)u_\Delta + B_2(\tau)d_\Delta +\eps_\Delta(x_\Delta).
\end{eqnarray}
By Taylor's theorem and the fundamental theorem of calculus, the error must satisfy,
\begin{eqnarray}
\begin{aligned}
\eps_{\Delta,i}(x_\Delta) = \frac{1}{2}x_\Delta^\top G(x_{\Delta,i}) x_\Delta \quad x_{\Delta,i} \in \tilde{\mathcal{B}},
\end{aligned}
\end{eqnarray}
where $G:\ssp \to \mathbb{R}$ is the Hessian of $\sys$ with respect to $x$ evaluated at a point, and $\tilde{\mathcal{B}}$ is a ball around $\tilde \tj$ with radius $\Vert x - \tilde \tj \Vert$. Moreover, the error is bounded by
\begin{eqnarray}
\begin{aligned}
|\eps_{\Delta,i}(x_\Delta) \vert &= \frac{1}{2}|x_\Delta^\top G(x_{\Delta,i}) x_\Delta| \leq \frac{1}{2}\Vert G(x_{\Delta,i})\Vert \: \Vert x_\Delta \Vert^2 \\ &\le \frac{1}{2} \max_{x \in \ftube}\Vert G(x - \tilde \tj)\Vert\:  \Vert x - \tilde \tj\Vert^2 \triangleq \delta_i^*
\end{aligned}
\label{TSerrdef}
\end{eqnarray}
where $\Vert G(\cdot)\Vert$ is the norm induced by the norm used for $\Vert x -\tilde \tj \Vert$. Since \cite{althoff2008reachability, althoff2011zonotope, chen2012taylor} provide convex polytopes $\fseto \supset \ftube$, \eqref{TSerrdef} may be solved on the order of milliseconds \cite{boyd2004convex}.

Hence, an envelope of the system may be solved with antagonistic error limited to
$\errset \triangleq \{ \Vert \text{Diag}(\maxerr)^{-1} x \Vert_\infty \le 1 \}$,
\begin{eqnarray}
\dot{x}= A(\tau) x + B_1(\tau)u + B_2(\tau)d + c(\tau) + \eps 
\end{eqnarray}
where $$c(\tau) \triangleq \sys(\tilde \tj, \tilde{u},\tilde{d}) - A(\tau) \tilde \tj(\tau) - B_1(\tau) \tilde u(\tau) - B_2(\tau)\tilde d(\tau).$$

Finally, we may solve the corresponding Hopf formula with the standard fundamental map approach, yielding
\begin{eqnarray}
\begin{aligned}
\ham_{\mathcal{Z}, \delta^*}(p,\tau) =  \ham_{\mathcal{Z}}(p,\tau) + \Vert\text{Diag}(\maxerr)p\Vert_1 + p^\top \Phi(\tau)c(\tau)
\end{aligned}
\end{eqnarray}
for $\ham_{\mathcal{Z}}(p,\tau)$ defined in (\ref{HamiltonianIndicator}).

\subsection{Multi-Agent Dubins Pursuit Evasion Benchmark Sets}

\end{document}